\documentclass{aa}
\usepackage{graphicx,epsfig,fancyhdr,rotate,amsmath,longtable}
\usepackage{natbib}
\bibpunct{(}{)}{;}{a}{}{,}
  \def\simge{\mathrel{\raise1.16pt\hbox{$>$}\kern-7.0pt
    \lower3.06pt\hbox{{$\scriptstyle \sim$}}}}           
  \def\simle{\mathrel{\raise1.16pt\hbox{$<$}\kern-7.0pt
    \lower3.06pt\hbox{{$\scriptstyle \sim$}}}}           
\begin{document}

      \title{Analysis of the Fe~{\sc x} and Fe~{\sc xiv} line width in the solar corona using LASCO-C1 spectral data}

      \subtitle{}

      \author{M. Mierla\inst{1,2} \and R. Schwenn\inst{1} \and L. Teriaca\inst{1} \and G. Stenborg\inst{3}
       \and  B. Podlipnik\inst{1}
             }


    \offprints{M. Mierla\\
    \email{m\_mierla@yahoo.co.uk}}
    \institute{Max-Planck-Institut f\"ur Sonnensystemforschung D-37191 Katlenburg-Lindau, Germany \and
      Astronomical Institute of the Romanian Academy, Bucharest, Romania \and
      Catholic University of America, Washington DC, 20064, USA}
    \date{Received: ; Accepted:}

    \abstract
    {}
    {The purpose of this paper is to analyze the variation in the line width with height in the inner corona
    (region above 1.1~$R_{\odot}$), by using the spectral data from
    LASCO-C1 aboard SOHO. We used data acquired at activity minimum (August - October 1996) and
    during the ascending phase of the solar cycle (March 1998).}
    {Series of images acquired at different wavelengths across the Fe~{\sc x}
    637.6~nm (red) and Fe~{\sc xiv} 530.3~nm (green) coronal lines by LASCO-C1
    allowed us to build radiance and width maps of the off-limb solar corona.}
    {In 1996, the line width of Fe~{\sc xiv} was roughly constant or increased with
    height up to around 1.3~$R_{\odot}$ and then it decreased. The Fe~{\sc x} line
    width increased with height up to the point where the spectra
    were too noisy to allow line width measurements (around 1.3~$R_{\odot}$). Fe~{\sc x} showed
    higher effective temperatures as compared with Fe~{\sc xiv}. In 1998 the line width of Fe~{\sc xiv}
    was roughly constant with height above the limb (no Fe~{\sc x} data available).}
    {}
   \date{Received 21 July 2007 / Accepted 20 november 2007}

   \titlerunning{Spectral line width analysis in the solar corona}

   \authorrunning{Mierla et al.}

   \maketitle


   \section{Introduction}

Spectroscopic observations of the solar corona in emission lines
help us in better understanding the dynamics of the solar corona.
Emission profiles from forbidden transitions of Fe~{\sc xiv}
(530.3~nm) and Fe~{\sc x} (637.6~nm) contain information on
physical parameters such as temperature, mass motion, and
turbulence. Coronal emission lines at visible wavelengths have
been observed using Fabry-Perot (FP) techniques or spectrographs
during total solar eclipses and with coronagraphs at high
altitudes (e.g., Singh 1985; Ichimoto et al. 1995; Hara and
Ichimoto 1999). This kind of observation is constrained by
scattering effects in the terrestrial atmosphere and rarely goes
up to 1.5~R$_{\odot}$ (Singh et al. 2006). The LASCO (Large Angle
Spectrometric Coronagraph, Brueckner et al. 1995) C1 aboard the
SOHO (Solar Heliospheric Observatory) spacecraft (Domingo et al.
1995), with its field of view (FOV) covering the solar corona
between 1.1 and 3~R$_{\odot}$, has provided unprecedented
observations of these spectral lines since its launch in December
1995 until June 1998. In the C1 design, a FP interferometer is
used as a narrow-passband, tunable filter. The instrument produces
a series of images that are separated in wavelength. This way, an
emission line can be sampled in wavelength, generating a complete
line profile simultaneously at each point in the FOV. The study of
the obtained line profiles can be invaluable in a) determining the
line widths and hence ion temperatures over the entire extent of
the corona, b) providing unambiguous evidence of mass flows in the
corona from the Doppler shift of the line (see Mierla et al.,
2005a).

 This paper concentrates on the line width, whose measurement can provide
 useful information concerning the ion temperatures, turbulent motions, and
 sub-resolution velocity fluctuations associated with magnetohydrodynamic waves in the
 solar corona.
 The intense forbidden lines of the solar corona are magnetic
 dipole transitions between levels of the ground configuration of
 highly ionized atoms. The emissivity of the green line is
 due to a forbidden transition of the Fe~{\sc xiv} ion at 530.3~nm
 (ground configuration: 3s$^2$3p and transition:
 $^2$P$_{3/2}$~$\--$~$^2$P$_{1/2}$). It peaks at a temperature of
 about 2~MK (Burgess and Seaton, 1964; Esser et al. 1995).
 The red coronal emission line at 637.4~nm (ground
 configuration: 3s$^2$3p$^5$ and transition:
 $^2$P$_{1/2}$~$\--$~$^2$P$_{3/2}$) from Fe~{\sc x} ions has its
 maximum emissivity at a temperature of around 1~MK.
 Below 1.2~R$_{\odot}$ the mode of excitation of Fe~{\sc x}
 637.4~nm line is predominantly collisional.
 Both collisional and radiative excitation (from the photospheric
 continuum radiation) are important for
 $1.2R_{\odot}<R<1.3R_{\odot}$, whereas radiative excitation becomes dominant
 beyond R=1.3, 1.4R$_{\odot}$ (Singh 1985, Raju \& Singh 1987).
 For Fe~{\sc xiv} the dominant excitation mechanism is collisional in the
 inner coronal regions (up to $\approx1.4~R_{\odot}$) (Raju et al. 1991).
 For a line profile of Gaussian shape, the Doppler width,
 $\Delta\lambda$, is related to the standard deviation, $\sigma$,
 and full width at half maximum (FWHM), through
 \begin {eqnarray}
 \Delta\lambda=\sigma\sqrt{2}={\rm FWHM}/(2\sqrt{\ln2}).\label{1}
 \end {eqnarray}
Assuming a Maxwellian velocity distribution of the photon-emitting
ions of mass $m_{\rm i}$, a relation between $\Delta\lambda$ and
the effective kinetic ion temperature $T_{\rm i}$ (comprising both
thermal and sub-resolution non-thermal motions) can be obtained:
\begin {eqnarray}
\Delta\lambda=(\lambda_{0}/c_{0})\sqrt{2kT_{\rm i}/m_{\rm i}}=
    (\lambda_{0}/c_{0})v_{\rm 1/e}
\label{2}
\end {eqnarray}
where {\it k} is the Boltzmann constant and $v_{\rm 1/e}$ is the
most likely speed along the line-of-sight (LOS). If the
instrumental profile can also be approximated by a Gaussian (of
Doppler width $\phi$), then we can write:

   \begin {eqnarray}
   \Delta\lambda=(\lambda_{0}/c_{0})\sqrt{2kT_{i}/m_{i}+{\phi}^2}\label{3}
   \end {eqnarray}

    \subsection{Spectroscopic observations of the solar corona}

    Based on Skylab data, Doschek et al. (1977), Nicolas et al.
    (1977), and Mariska et al. (1978) have shown that the line width increases
    with height above the limb. These earlier observations were restricted to
    emission lines emitted by chromospheric or transition region ions formed at
    temperatures lower than 2.2$\cdot10^{5}$~K. Hassler et al. (1990), analyzing Mg~{\sc x} (60.9~nm and 62.5~nm),
    Fe~{\sc xii} (124.2~nm), O~{\sc v} (62.9~nm), and N~{\sc v}
    (123.8~nm and 124.2~nm) line profiles from the off-limb quiet corona, have
    seen an increase in the line widths with altitude up to around 0.2~$R_{\odot}$ above the limb.

    Wilhelm et al. (2004) used SUMER spectra of the Mg~{\sc x} (62.5~nm) line
    to find that the line width broadens from around 8.2~pm to
    around 9.5~pm between the limb and 220~Mm (0.33~R$_{\odot}$)
    above the limb in the equatorial corona. Harrison et al. (2002),
    using CDS data, report the Mg~{\sc x} (62.5~nm) line
    narrowing with height at altitudes above 50000 km in the quiet
    near-equatorial solar corona. The authors
    suggest that emission line broadening, at lower altitudes, is due to
    the outward propagation of undamped Alfv\'{e}n waves,
    in open magnetic field regions with decreasing density with altitude. The
    narrowing at higher altitudes is interpreted as further evidence
    of coronal waves activity, but in closed magnetic field regions.
    Still in equatorial regions, Doyle et al. (1998) find an increase in the
    Si~{\sc viii} line width with height above the limb, from about
    0.026~nm at the limb to 0.028~nm at 25000~km (0.04~R$_{\odot}$)
    above the limb.
    Wilhelm et al. (2005) analyze spectra from combined CDS and
    SUMER observations of the relatively quiet equatorial corona above
    a small prominence and find no or very slight increases with altitude in
    the widths of emission lines from ions of several different elements.
    In polar regions, Banerjee et al. (1998) and Doyle et al. (1999)
    used SUMER spectra of Si~{\sc viii} lines to observe an increase in the
    non-thermal velocity from 27~km~$s^{-1}$ at 27 arcsec above the limb to more
    than 46~km~$s^{-1}$ at 250 arcsec (0.25~R$_{\odot}$) above the limb.
    An increase in the Mg~{\sc x} 62.5~nm width with height in polar
    coronal holes was also found by Wilhelm et al. (2004) in SUMER spectra.
    On the other hand, O'Shea et al. (2005), using Mg~{\sc x} 60.97~nm and
    62.49~nm spectra from CDS, observed a decrease of the line width
    with height in coronal holes. They  suggest that the line widths start
    to show a decrease in their values at the location, around 0.15~R$_{\odot}$
    above limb, where the dominant excitation changes from
    collisional to radiative. The decrease in the line width with
    height in the solar corona was explained by Zaqarashvili et
    al. (2006) by resonant energy conversion from Alfv\'{e}n to
    acoustic waves near the region of the corona where the plasma
    $\beta$ approaches unity.

    Chandrasekhar et al. (1991) have observed from eclipse data that
    the width of the green line increases up to 1.2~R$_{\odot}$ and then decreases with height,
    while for the red line it increases with the radial
    distance up to 1.2~R$_{\odot}$ from the Sun center.
    In coronal structures, the width of the red line
    increases with height above the limb at a rate between 0.05 and 0.26~pm/arcsec,
    whereas the green line width in the same region
    decreases with height at a rate of 0.12 to 0.34 pm/arcsec
    (Singh et al. 1999, 2002). The authors explain this behavior as
    resulting from the increased mixing of hotter and cooler plasma with height. The mixed
    plasma at large heights will have an average temperature lower than that
    of the green line plasma and higher than the red line plasma at
    the bottom of the corona.
    Singh et al. (2003a) have observed an increase in the Fe~{\sc x}
    637.4~nm and Fe~{\sc xiii} 1074.7~nm, 1079.8~nm line widths at the rates of
    0.124~pm/arcsec and 0.029 pm/arcsec, respectively, and a decrease
    in Fe~{\sc xiv} 530.3~nm line width at a rate of -~0.066~pm/arcsec. They
    suggest that, if the ionization temperature of the species is
    higher than 1.6~MK, then the line width is decreasing with
    height, and if the ionization temperature is lower than
    1.6~MK, then the line width is increasing with height above
    the limb. Singh et al. (2003b) have observed that in steady coronal
    structures the FWHM of the red line increases with height above the
    limb at a rate of around 0.102~pm/arcsec and the FWHM of the
    \begin{figure*}[!th]
      \centering
      \includegraphics[width=.35\textwidth,type=eps,ext=.eps,read=.eps]{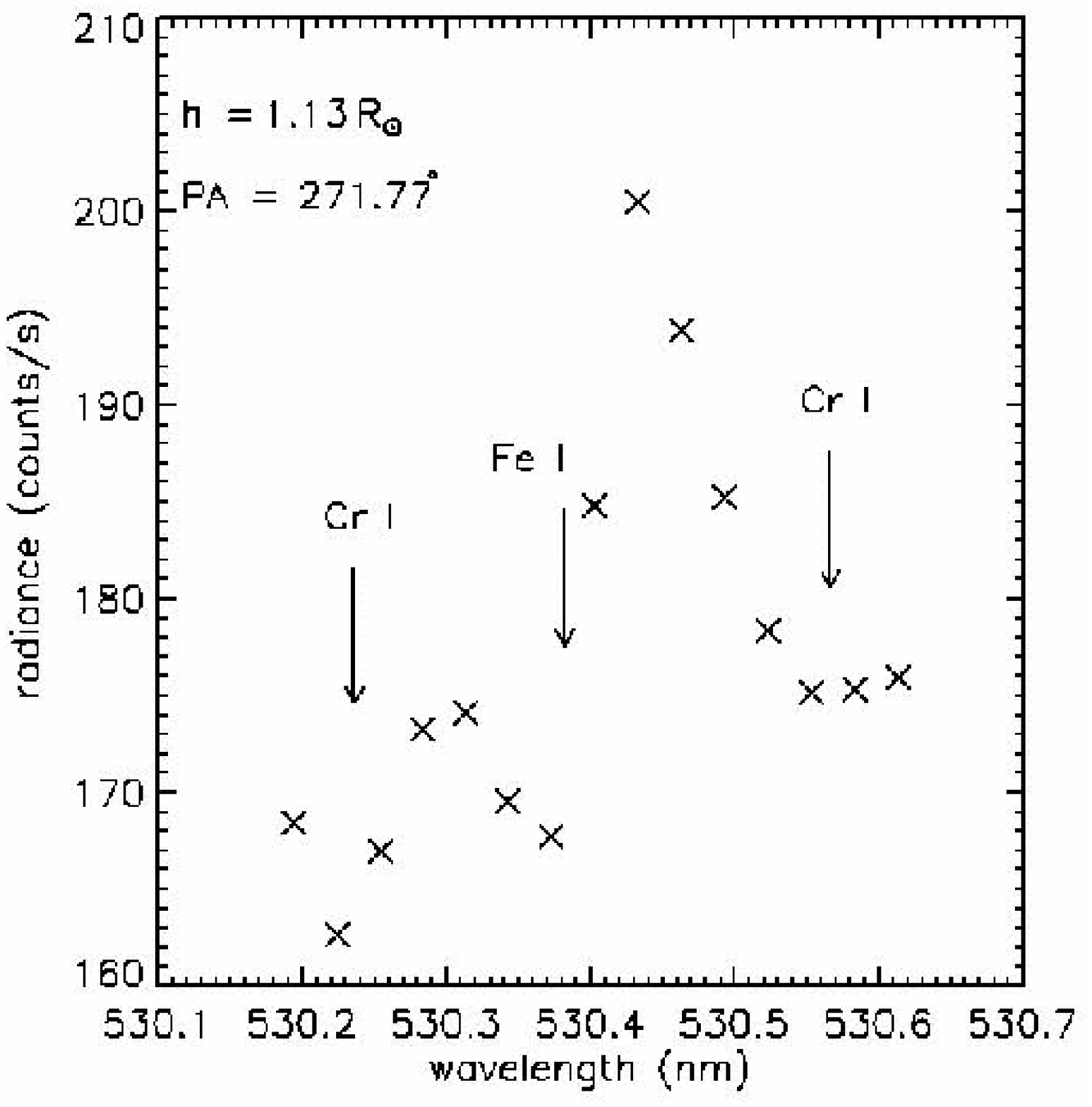}
      \includegraphics[width=.35\textwidth,type=eps,ext=.eps,read=.eps]{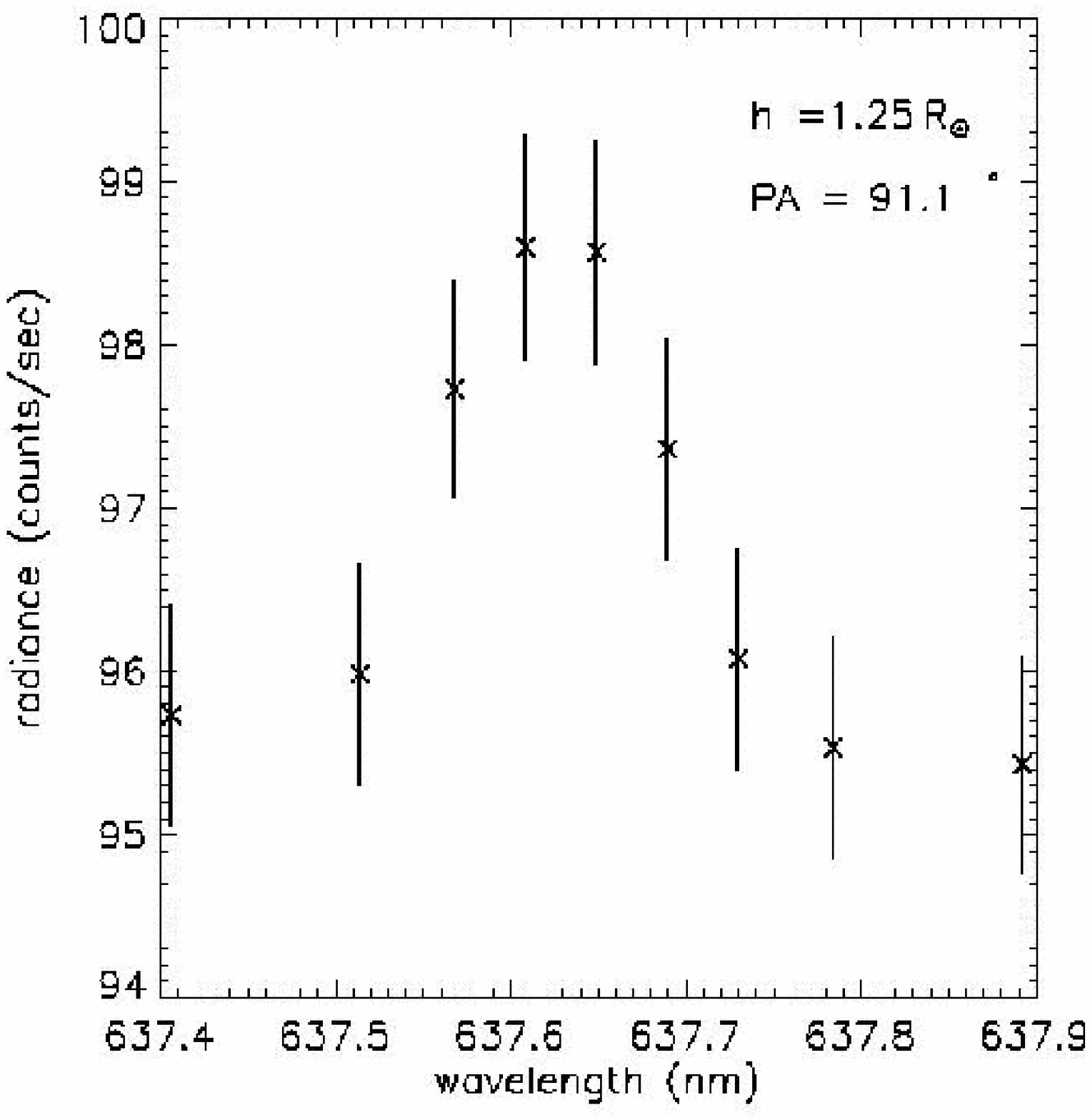}
      \caption{Left panel: Fe~{\sc xiv} profile of the coronal region (PA 271
      and d=1.13$R_{\odot}$), as observed by LASCO-C1 on 28 March 1998.
      The observed profile is affected by the superimposition of
      the instrumental straylight photospheric spectrum comprising 3 nearby
      absorption line profiles (Cr~{\sc i} $\approx$530.23~nm,
      Fe~{\sc i} $\approx$530.37~nm, and Cr~{\sc i} $\approx$530.56~nm).
      Right panel: Fe~{\sc x} profile of a coronal region (PA 91 and
      d=1.25~R$_{\odot}$), as observed by LASCO-C1 on 20 August 1996.
      The errorbars were computed using Poisson distribution for the
      photoelectrons (one count corresponds to 13 photoelectrons).
      The position angle (PA) is measured counterclockwise from the north pole.}
    \label{c1profiles}
   \end{figure*}
    Fe~{\sc xi} 789.2~nm line also increases with height but at a lower average
    value: 0.055~pm/arcsec. Singh et al. (2004) analyzed the data
    of Fe~{\sc x} (637.4~nm), Fe~{\sc xi} (789.2~nm), Fe~{\sc xiii}
    (1074.7~nm), and Fe~{\sc xiv} (530.3~nm) recorded with the coronagraph
    at Norikura Solar Observatory. They noticed that the FWHM of
    the Fe~{\sc x} line increased with height in all coronal structures
    except for two, where the FWHM of the Fe~{\sc x} line
    decreased with height. The FWHM of the Fe~{\sc xiv} line decreased
    with height in almost all of the structures. From further high spectral and spatial resolution observations
    at Norikura, Singh et al. (2006) observe that a) the FWHM of the
    Fe~{\sc xiv} line decreases up to around 300 arcsec (0.3~R$_{\odot}$)
    above the limb and then it remains more or less the same up to 500~arcsec
    (0.5~R$_{\odot}$), and b) the FWHM of Fe~{\sc x} line increases up to
    250~arcsec (0.25~R$_{\odot}$) and subsequently remains
    constant.

    \section{The instrument and the observations}

    The LASCO instrument consists of three
    individual coronagraphs with nested FOVs. The C1
    coronagraph observes the solar corona from 1.1 to 3.0~$R_{\odot}$
    and contains a FP interferometer that allows imaging the corona in
    different emission lines. The FP has a bandpass (FWHM) of 0.07~nm
    and a tunable range of 1~nm. Blocking filters, which are used as
    order sorters, have a free spectral range (interorder separation)
    of 3.5~nm. The detector is a CCD camera with a nominal spatial
    resolution of 11.2 arcsec (pixel size of 5.6 arcsec) (Brueckner et al. 1995).

    The data analyzed here are images taken in the lines of Fe~{\sc
    xiv} at 530.3~nm (green coronal line) and of Fe~{\sc x} at
    637.6~nm (red coronal line) in two periods of time: 1996 (minimum
    of solar activity) and 1998 (on the rising phase of the solar
    cycle). The green line data of 1998 consist of 47
    sets (from 28 to 30 March) each consisting of two
    scans of both the west and east limbs of the Sun. Each scan
    consists of 15 images at wavelengths ranging from 530.20~nm to
    530.65~nm in steps of 0.03~nm, along with 2 off-line images at
    wavelengths of 531.14~nm and 529.99~nm (the wavelengths are given
    relative to vacuum). The images were taken with a cadence of
    1~min, the exposure time being 25~s. Thus, spectra are recorded in
    about 15~min. The data of 1996 were taken between 1 August
    and 22 October in both the green and red lines. Each set
    consists of 9 spectral on-line images, ranging from 530.32~nm to
    530.62~nm for Fe~{\sc xiv} data and from 637.43~nm to 637.92 nm
    for Fe~{\sc x} data, along with one off-line image
    taken at a wavelength of 531.14~nm (green) and 638.10~nm (red). In
    this period 1-2 data scans were taken per day, with an image cadence of 3
    min and an exposure time of 25~s (green) and 16~s (red). In
    total, 97 (green) and 76 (red) data sets were recorded.
   \begin{figure*}[!t]
   \centering
   \includegraphics[width=.3\textwidth,type=eps,ext=.eps,read=.eps]{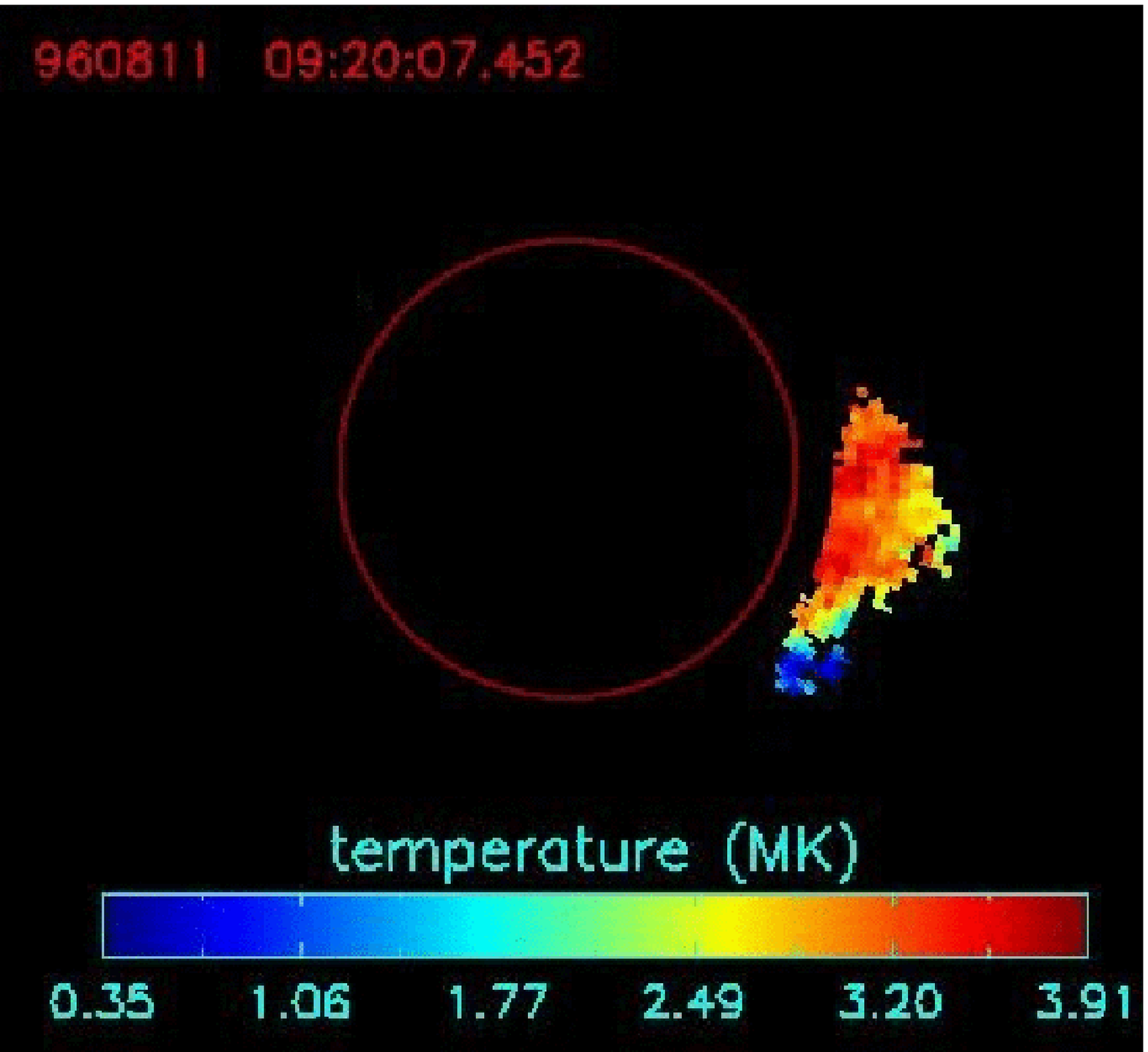}
   \includegraphics[width=.3\textwidth,type=eps,ext=.eps,read=.eps]{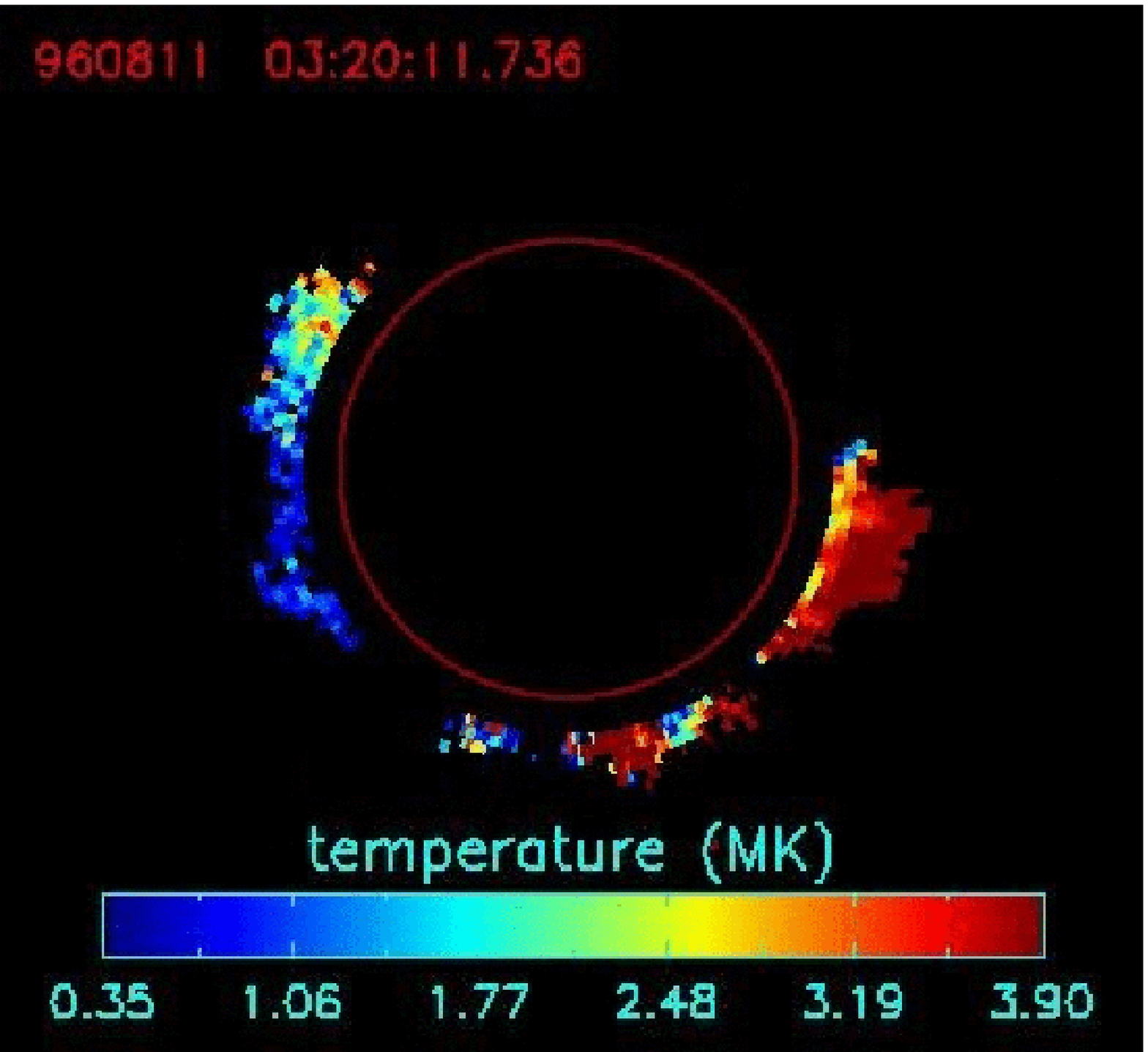}
   \caption{Example of Fe~{\sc xiv} (left panel) and Fe~{\sc x} (right panel)
    effective temperature maps for the data recorded on 11 August 1996.
    Larger widths (higher ion temperatures) are represented in red.}
   \label{c1widthmap}
   \end{figure*}
    \section{Data reduction}
    \begin{figure*}[!th]
    \centering
    \includegraphics[width=.35\textwidth,type=eps,ext=.eps,read=.eps]{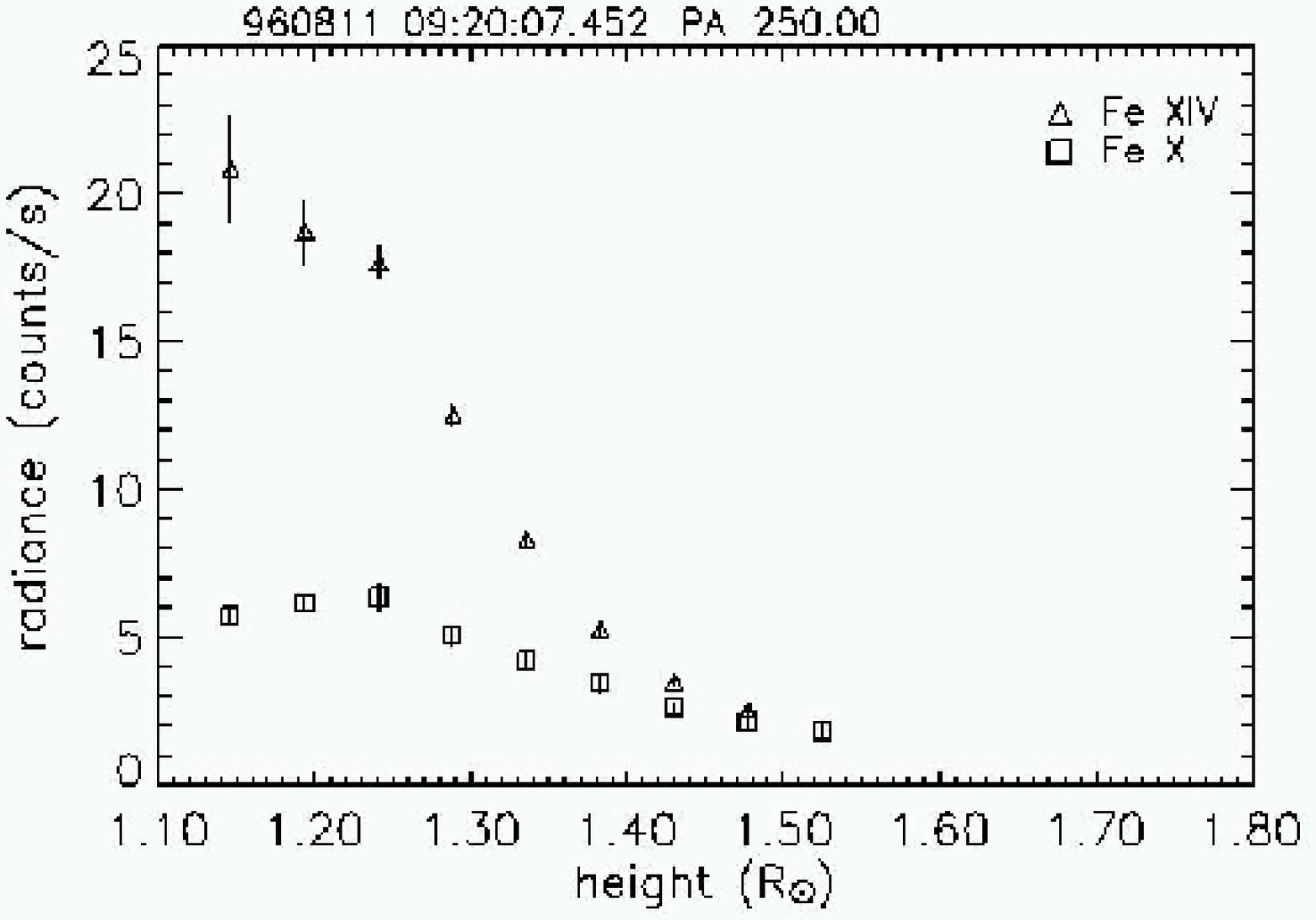}
    \includegraphics[width=.35\textwidth,type=eps,ext=.eps,read=.eps]{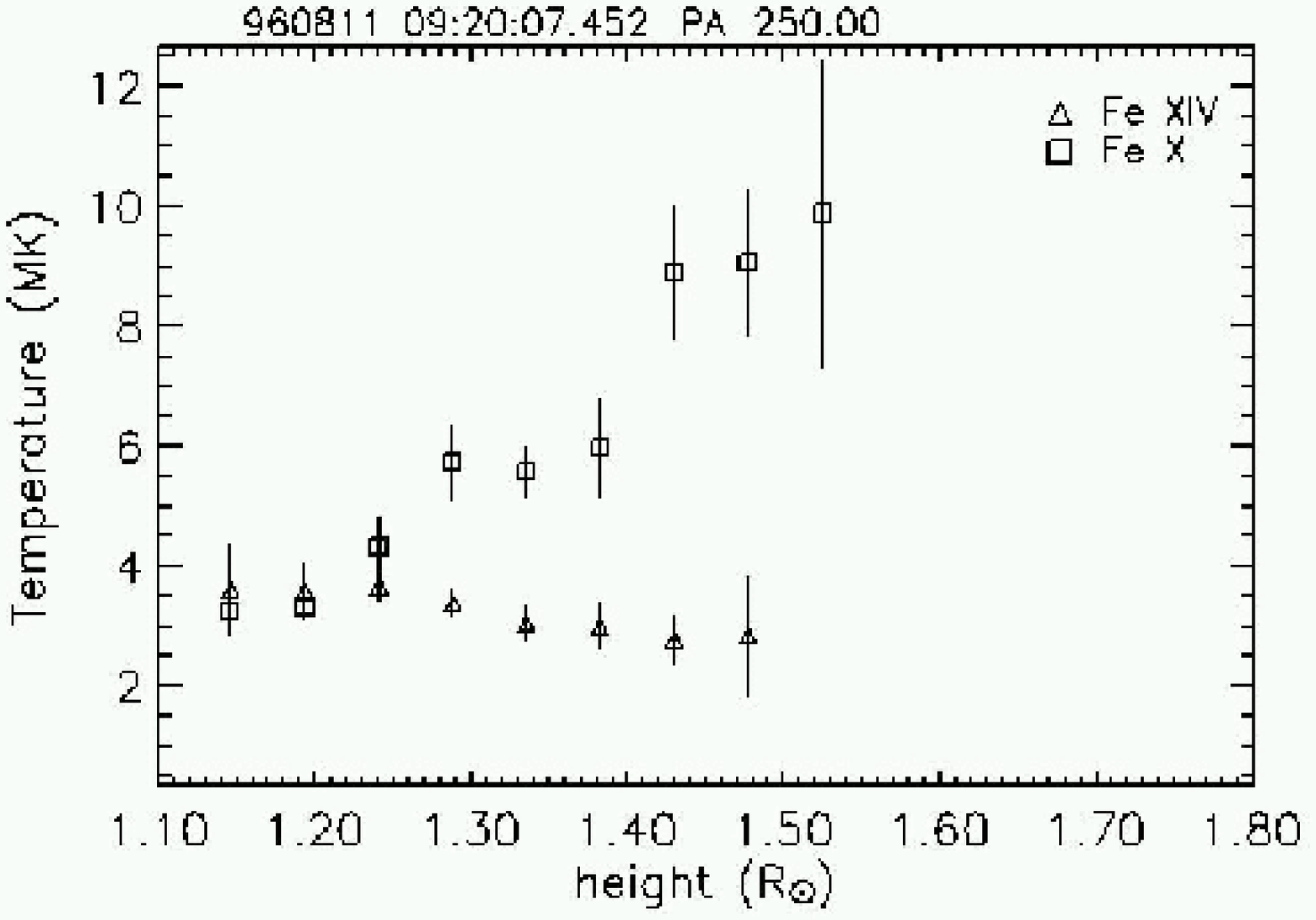}
    \includegraphics[width=.35\textwidth,type=eps,ext=.eps,read=.eps]{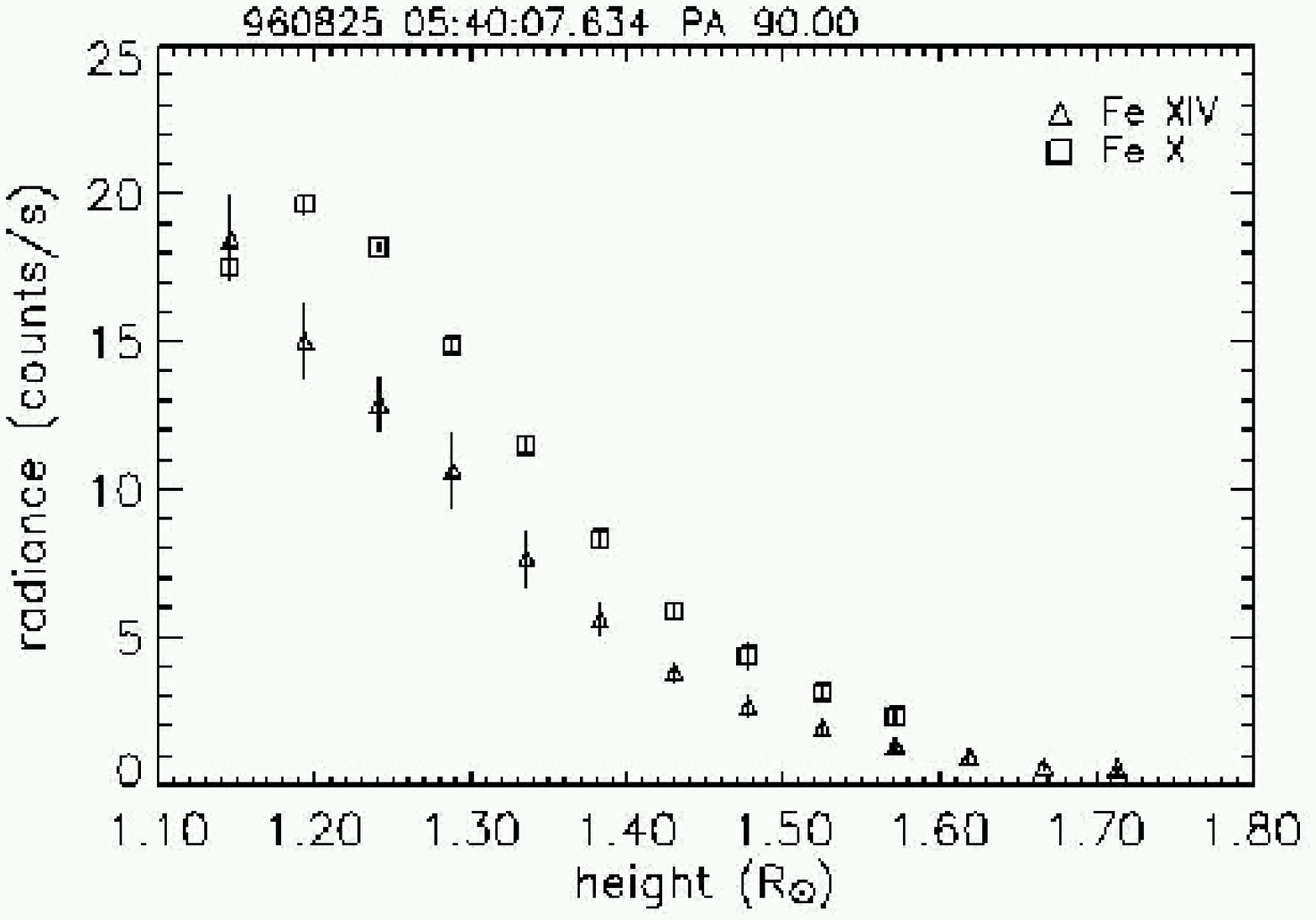}
    \includegraphics[width=.35\textwidth,type=eps,ext=.eps,read=.eps]{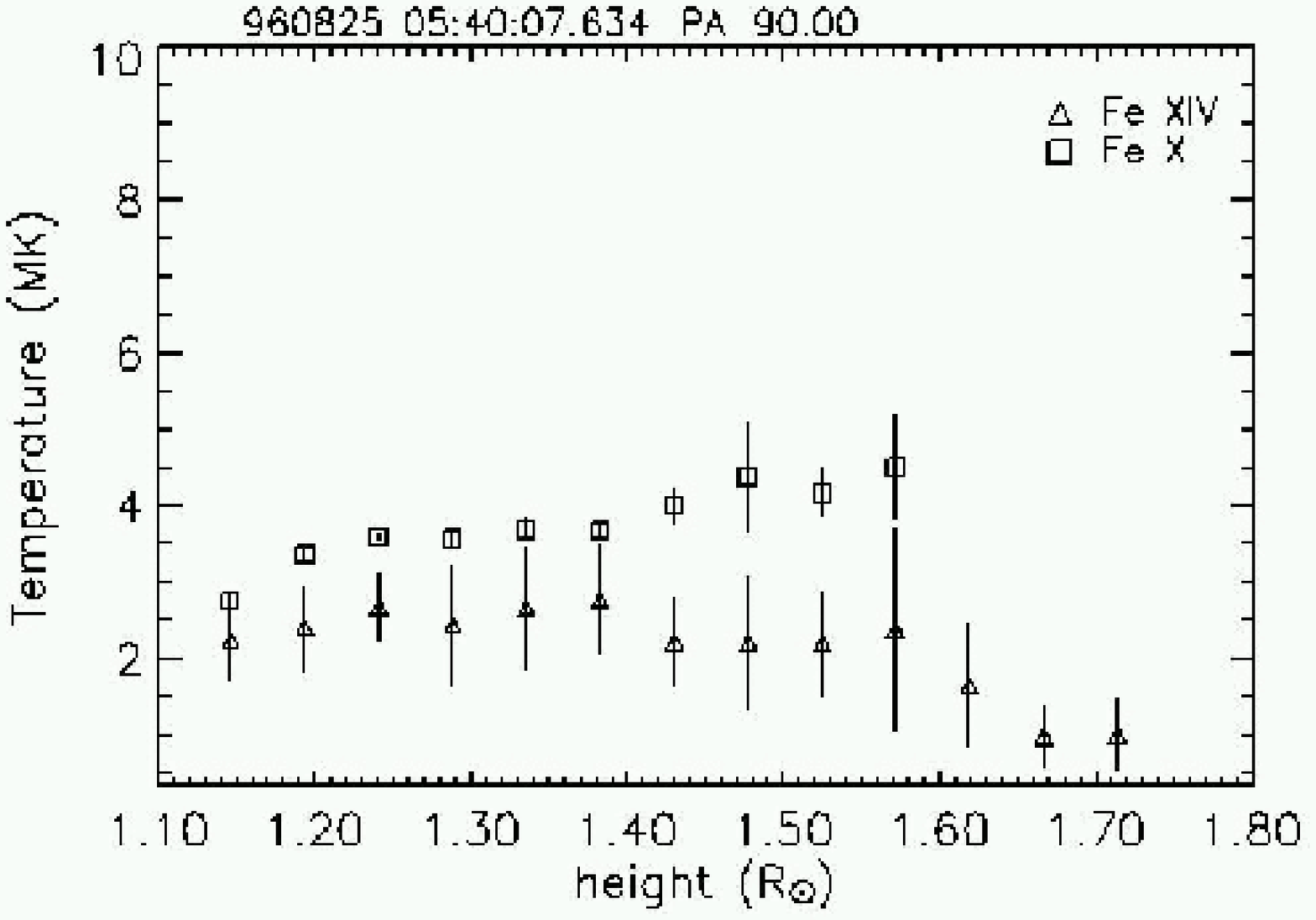}
    \includegraphics[width=.35\textwidth,type=eps,ext=.eps,read=.eps]{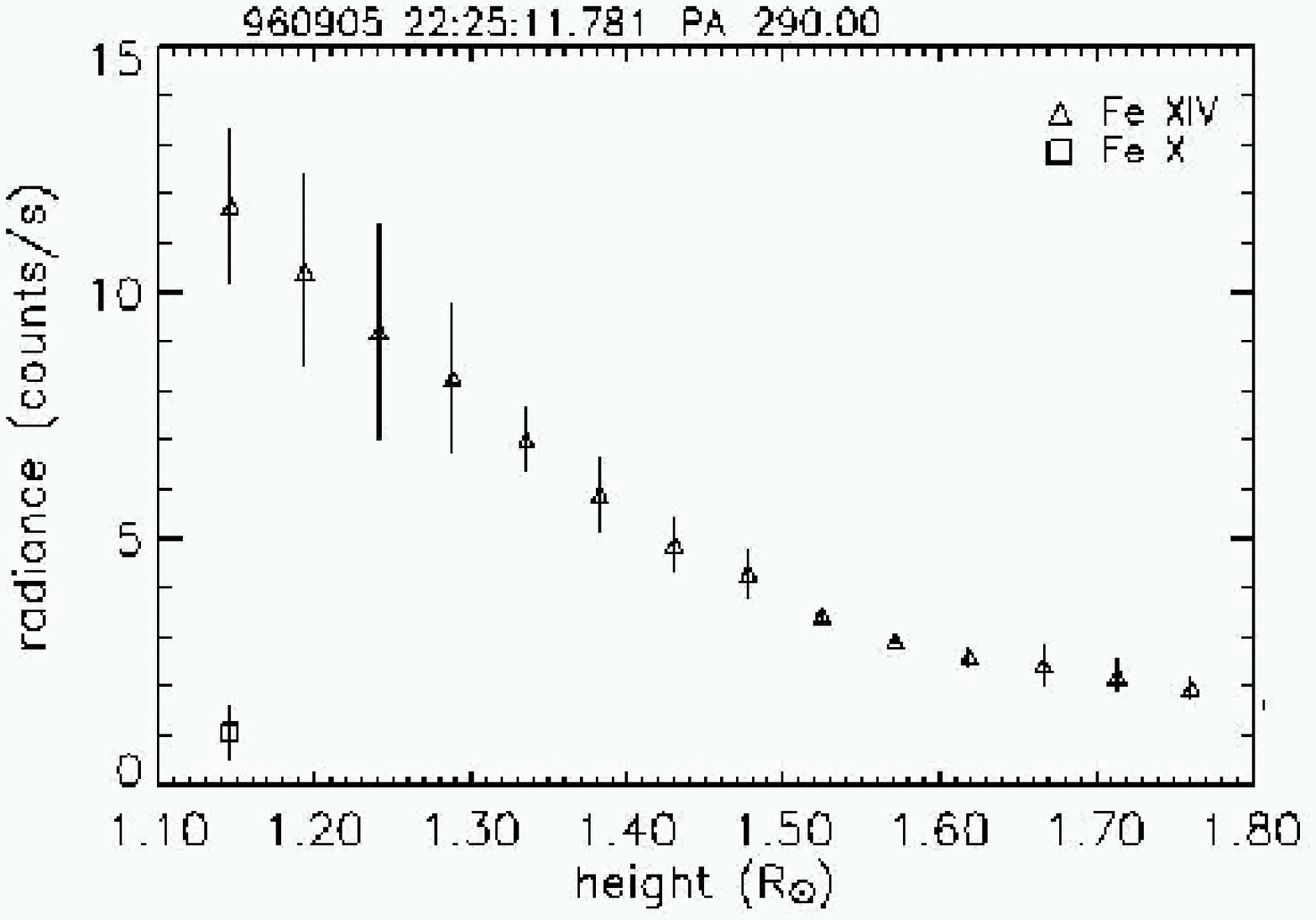}
    \includegraphics[width=.35\textwidth,type=eps,ext=.eps,read=.eps]{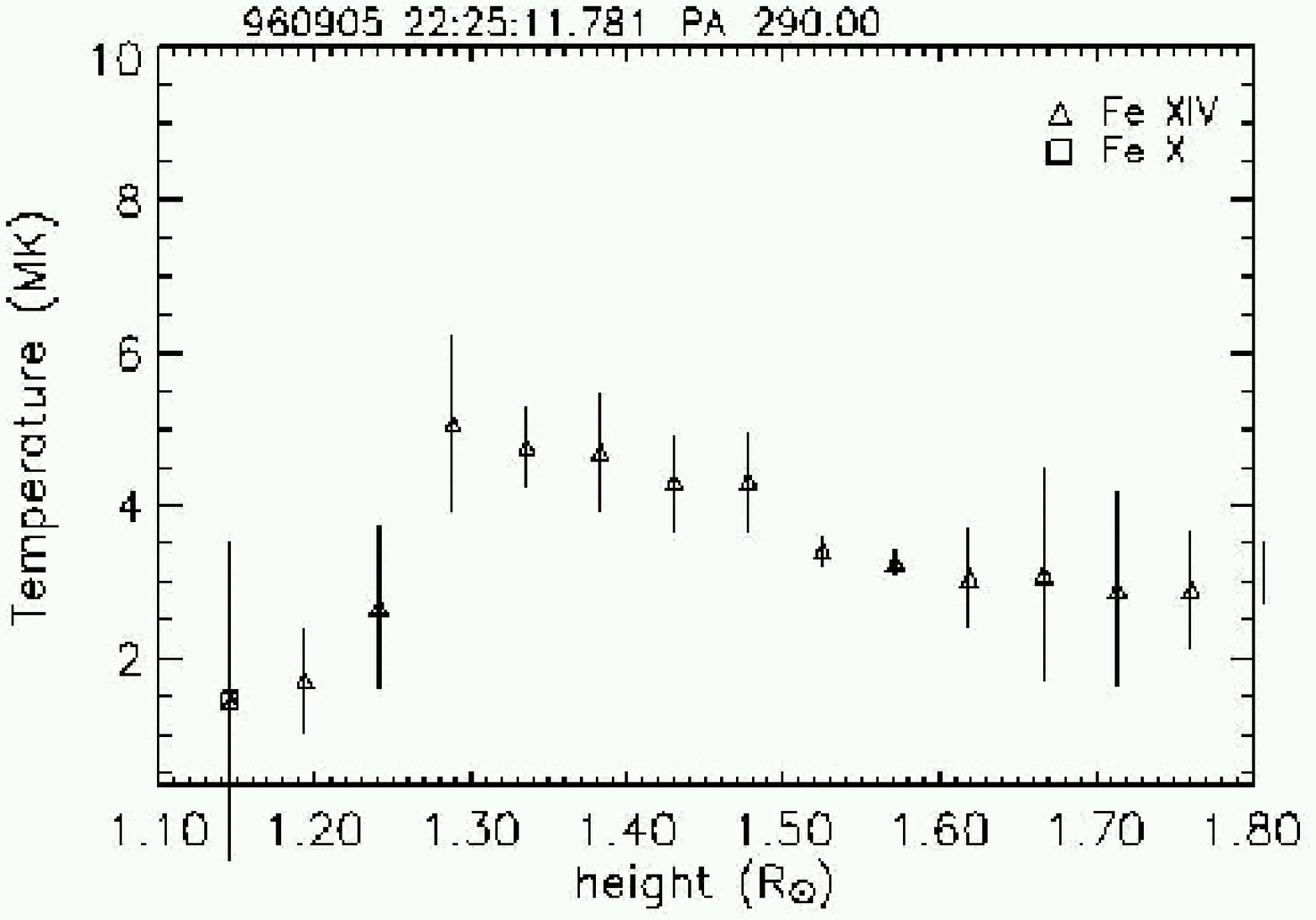}
    \caption{Example of radial plots
    of the emission line radiance (left) and temperature (right panels) taken at the indicated polar angles
    on 11 August (upper panels), 25 August (middle panels), and 5 September (lower panels).}
    \label{plotsmin11aug}
    \end{figure*}
    A detailed description of the data reduction is given in Mierla et al.
    (2005a). To summarize, the following steps are applied. (1) In bias
    subtraction, the bias was added electronically on board in
    order to avoid negative values of the signal count rates. (2)
    For cosmic-ray hits removal, the corrupted values due to high
    energetic particles were removed by using a median filter
    technique. (3) For exposure time correction, the correction is only applied
    to Fe~{\sc x} data because of the smaller fraction (compared
    with the Fe~{\sc xiv} data) of count-rates due to Fe~{\sc x}
    photons with respect to the total signal (see
    Fig.~\ref{c1profiles}). The emission lines are
    sitting on top of an intense continuum spectrum (mainly
    instrumental straylight). Thus, any small inaccuracy in the
    exposure time could affect the emission line profile when a model
    (errorless) background is subtracted. (4) For relative calibration of the
    wavelengths in the FOV of each image, the peak transmission
    wavelength of the FP depends on the angle of incidence that is
    equivalent to different spectral positions over the detector
    plane. (5) Absolute wavelength calibration using the main
    Fraunhofer line at 530.37~nm (Fe~{\sc i}) which was observed as scattered
    light in the instrument (Mierla et al. 2005a, 2005b). This does not apply
    to Fe~{\sc x} data, because the peak of the line is far from any
    absorption line.

   Line profiles are built by taking the radiance at a given
   pixel, or an ensemble of adjacent pixels (i.e., superpixel), of
   all images in a set as a function of the corresponding wavelengths
   (Fig.~\ref{c1profiles}). The observed spectral profile consists of
   line emission from ions in the corona, photospheric light
   scattered at free electrons in the corona (``continuum corona''), and
   instrumental scattered light. The last component contains the
   well-known Fraunhofer lines, some of which are partially superimposed on the Fe~{\sc xiv} emission
   line profile (see left panel of Fig.~\ref{c1profiles}). In the
   case of the Fe~{\sc x} emission line, there are no relevant absorption lines
   at nearby wevelengths and the profile is represented well by a Gaussian (see
   right panel of Fig.~\ref{c1profiles}).
   In order to improve the signal to noise ratio for the data in 1996 we created
   superpixels of four pixels in the radial direction and one degree
   (i.e., $\approx$4~pixels at 1.1~R$_{\odot}$,
   $\approx$ 10~pixels at 3~R$_{\odot}$)
   in the azimuthal direction. The obtained line profiles were finally fitted.
   For the data in 1998, we have enough signal to fit the
   data at each pixel of the images.
    \begin{figure*}[!th]
    \centering
    \includegraphics[width=.34\textwidth, type=eps,ext=.eps,read=.eps]{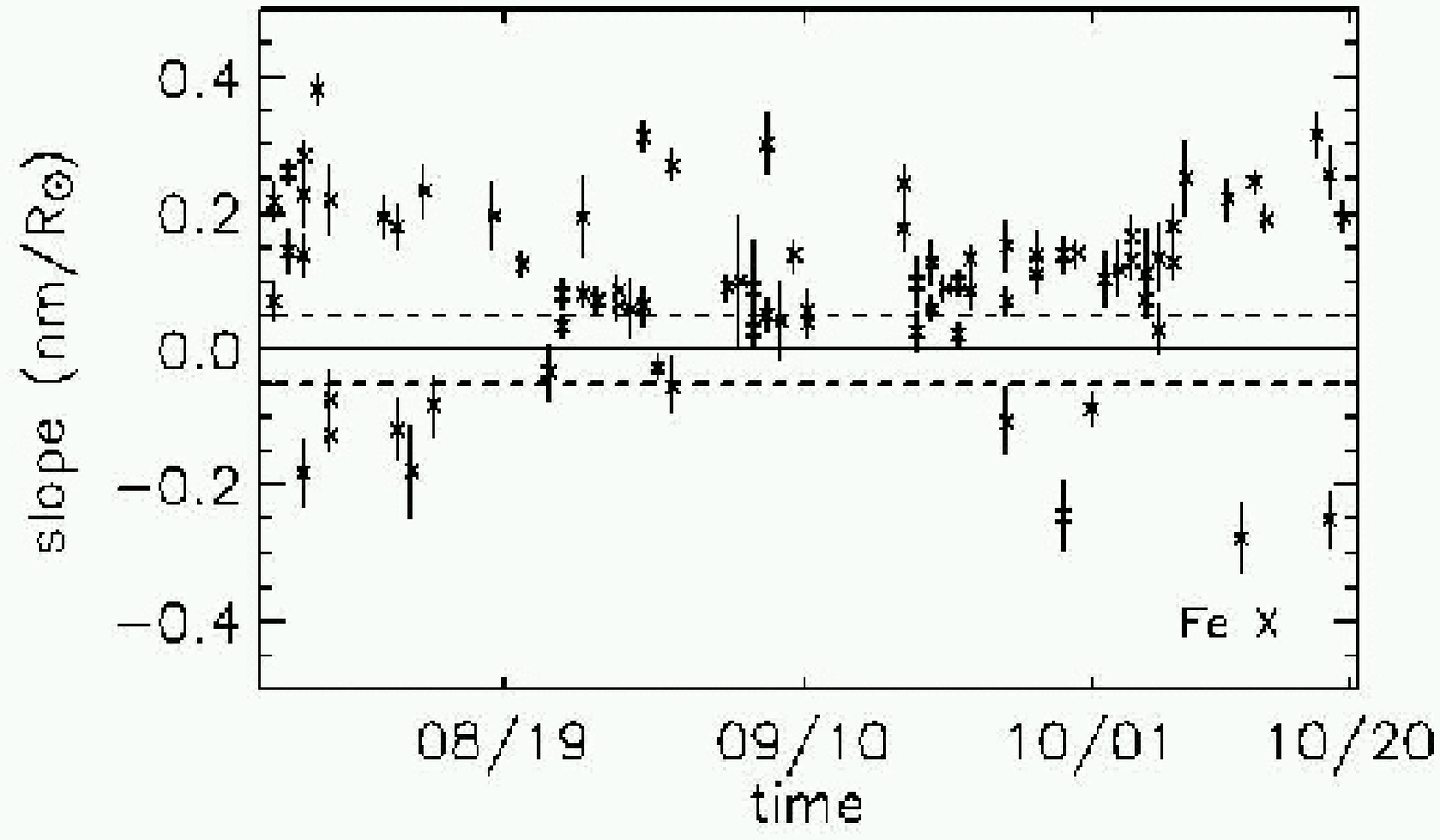}
    \includegraphics[width=.34\textwidth, type=eps,ext=.eps,read=.eps]{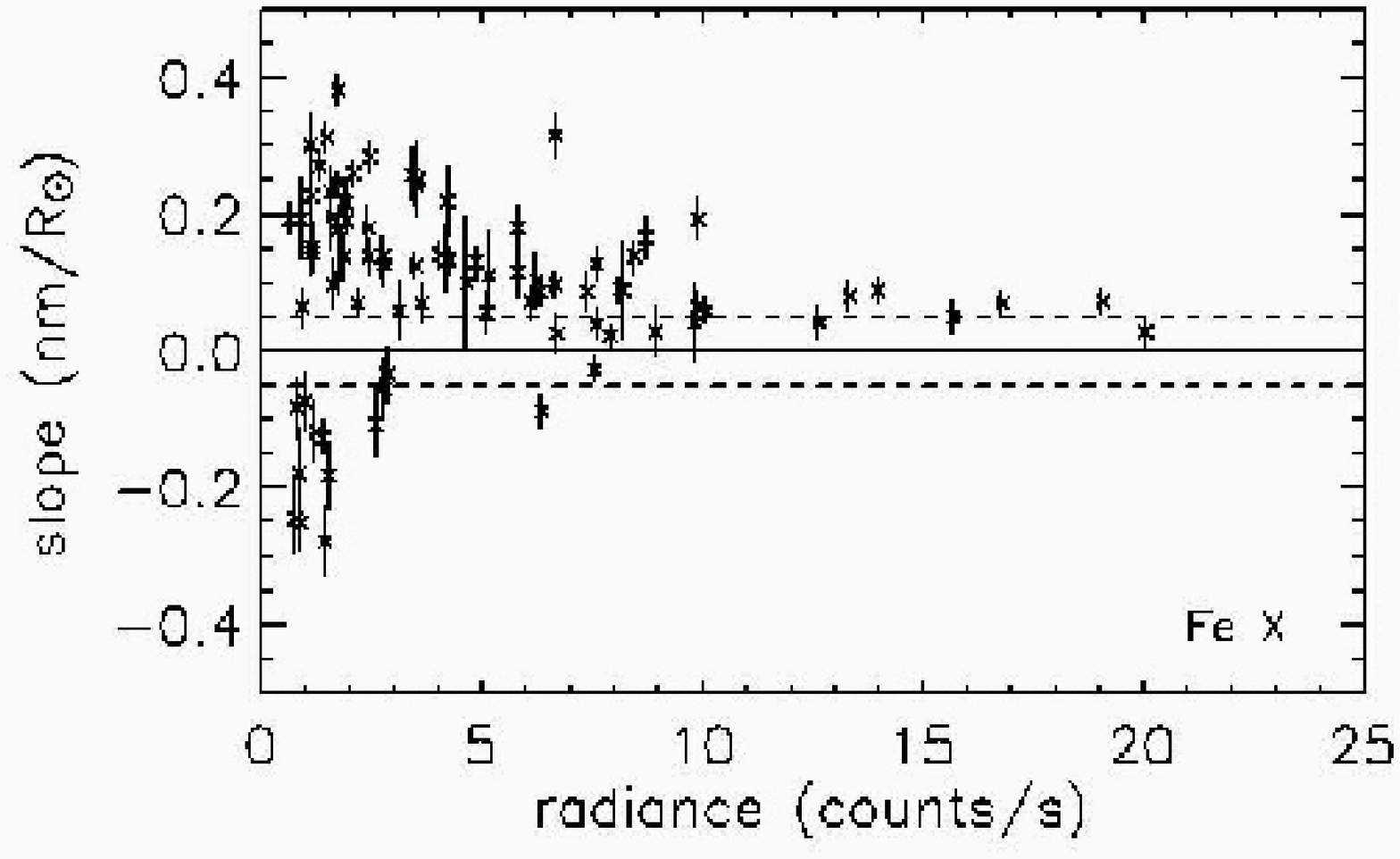}
    \caption{Left panel: The Fe~{\sc x} slope for the data in 1996, taken at the PA: 70,
    90, 110, 250, 270, 290. The continuous horizontal line indicates
    a slope of zero, while the two dotted lines indicate slopes of
    $\pm0.05$~nm/R$_{\odot}$. The X-axis represents the date in format mm/dd. Right panel: The variation
    in the Fe~{\sc x} slope with the radiance at
    1.2~R$_{\odot}$ for the same data as in left panel.}
    \label{fexslope96}
    \end{figure*}
    \begin{figure*}[th!]
    \centering
    \includegraphics[width=.34\textwidth, type=eps,ext=.eps,read=.eps]{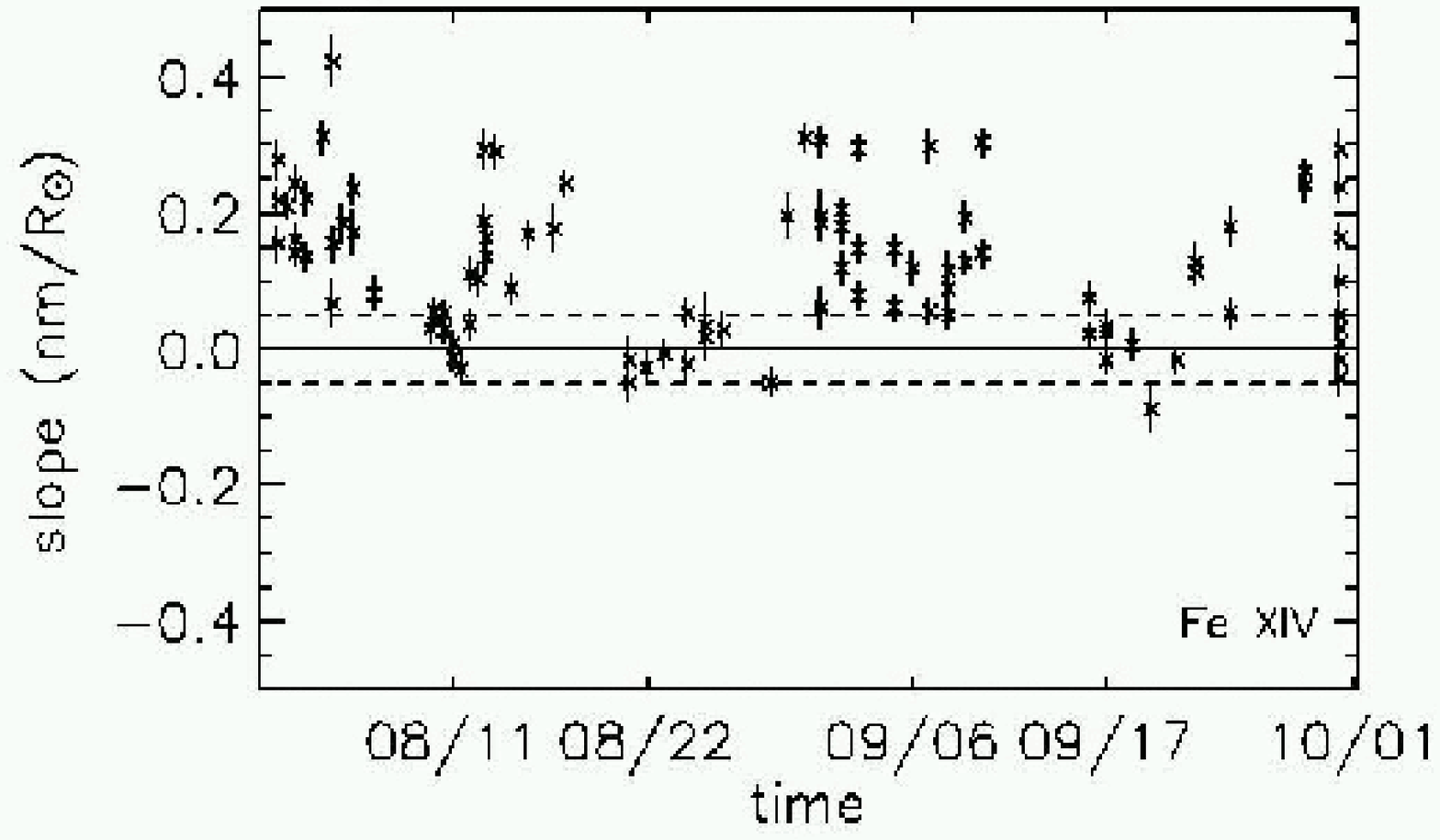}
    \includegraphics[width=.34\textwidth, type=eps,ext=.eps,read=.eps]{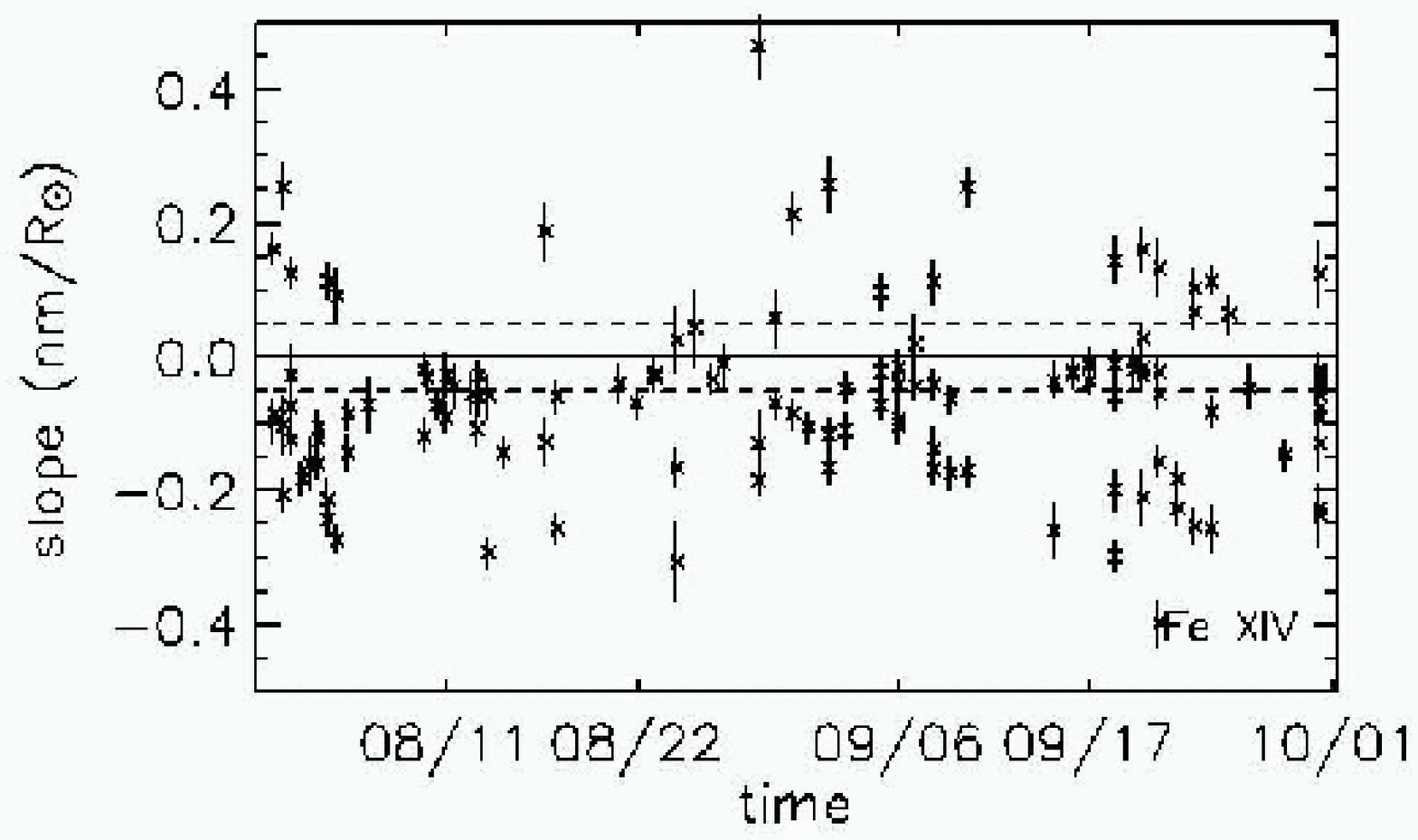}
    \includegraphics[width=.34\textwidth, type=eps,ext=.eps,read=.eps]{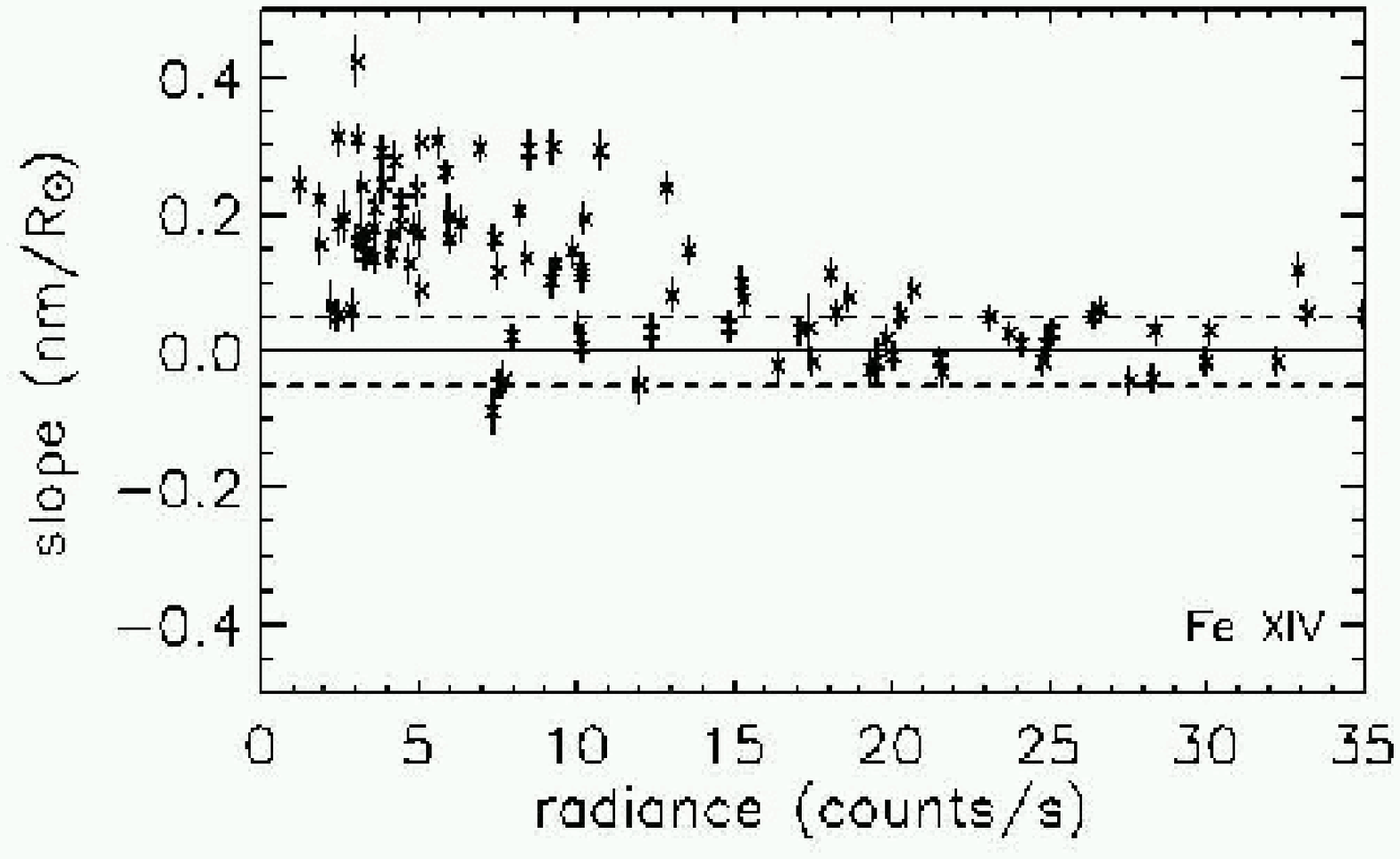}
    \includegraphics[width=.34\textwidth, type=eps,ext=.eps,read=.eps]{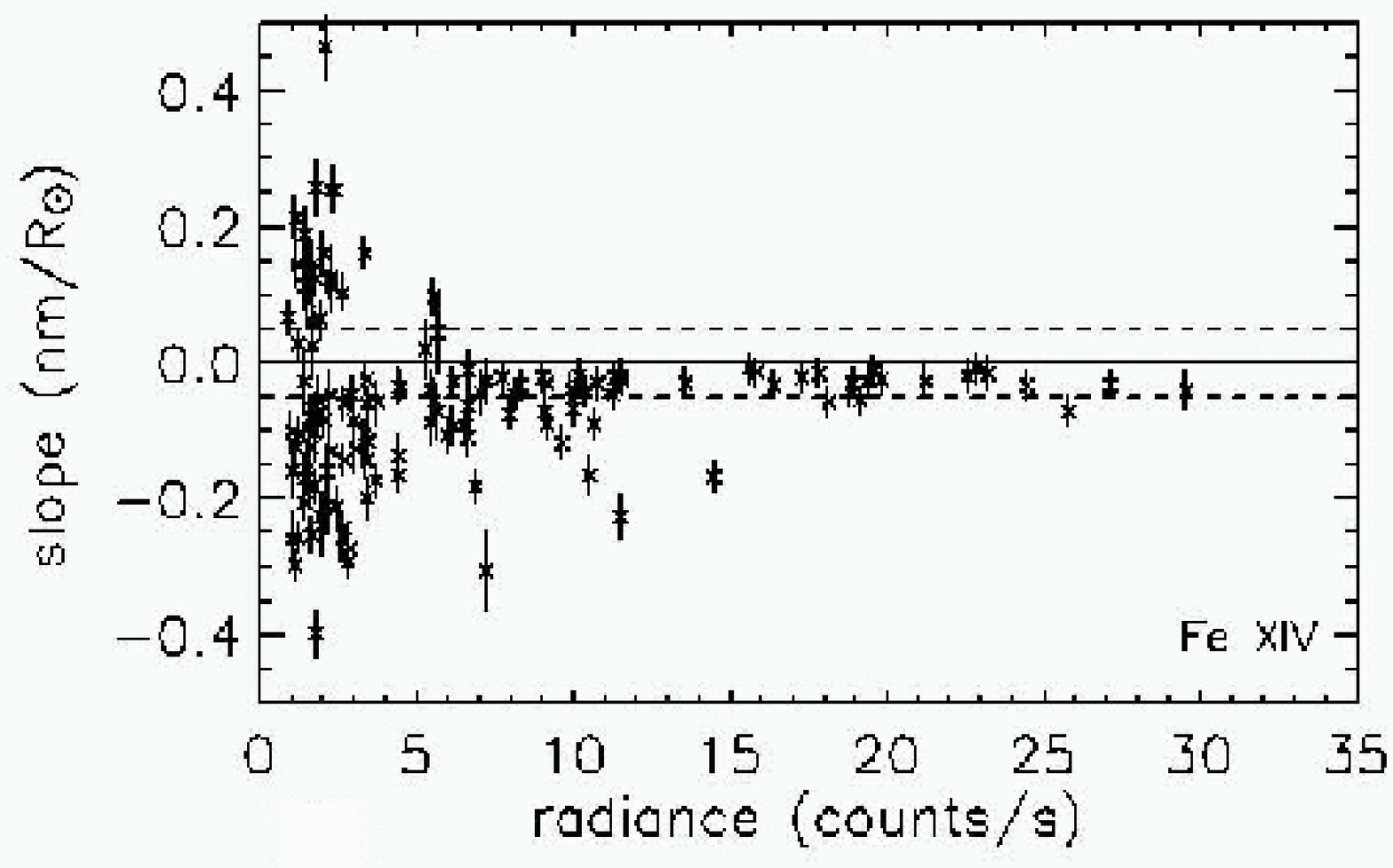}
    \caption{Upper panels: The variation in the Fe~{\sc xiv} slope with time for the data in
    1996. The upper-left panel shows the slopes for the data between
    1.1 and 1.3~R$_{\odot}$ and the upper-right panel data between 1.3 and 1.5~R$_{\odot}$.
    The slopes are calculated for the data taken at the PA: 70,
    90, 110, 250, 270, 290. On the X-axis is represented the date in format mm/dd. Lower panels:
    The variation of Fe~{\sc xiv} slope with the radiance at 1.2~R$_{\odot}$ (lower-left panel)
    and 1.4~R$_{\odot}$ (lower-right panel) for the same data as in the above panels.}
    \label{fexivslope96}
    \end{figure*}
   The fitting procedure for Fe~{\sc xiv} data is described in detail
   by Mierla et al. (2005a). For the Fe~{\sc x} data, a fit with one
   Gaussian and a constant background is performed. The FWHM maps are
   built by replacing each pixel (or superpixel) with the line FWHM
   values. The line was corrected for the instrumental profile
   (0.06~nm in the case of Fe~{\sc xiv} line and 0.08~nm in the case
   of Fe~{\sc x}; see Brueckner et al. 1995). For illustration, we
   show one example from data taken on 11 August 1996
   (Fig.~\ref{c1widthmap}). An FWHM of 0.07~nm for Fe~{\sc xiv}
   corresponds to an effective temperature of around 2~MK, and an FWHM of 0.06~nm
   for Fe~{\sc x} is equivalent to 1~MK. The regions where the fit
   could not be performed due to the weak line radiances are
   represented in black. In general, we can
   see the coronal features up to around 1.8~R$_{\odot}$ with our spectral
   data, and reliable measurements of line widths can be performed up to
   1.6~R$_{\odot}$ for Fe~{\sc xiv} data and
   up to 1.3~R$_{\odot}$ for Fe~{\sc x} data.
    \section{Data interpretation}

    \subsection{The line widths and the solar corona at activity minimum}
    By analyzing the effective temperature maps we found
    that the effective ion temperatures are higher in the red corona
    than in the green corona, although the Fe~{\sc x} line has a
    lower formation temperature. This may mean that the amount of
    turbulence is higher in the cooler plasma. However, high
    in the corona, ions and electrons are very likely not in
    equilibrium, and the formation temperature (kinetic temperature of
    the electrons determining the ionization fraction) may not have
    anything to do with the ion temperature. In this context our
    data may indicate a preferential heating of the less ionized
    species (lower $q/m$) as required by ion-ciclotron dissipation
    of Alfv\'{e}n waves (Tu et al. 1998).
    From the data taken on 11 and 25 August 1996
    (upper and middle panels of Fig.~\ref{plotsmin11aug}),
    at polar angles (PA) 250 and 90, respectively, we
    notice that the Fe~{\sc xiv} line width is roughly constant up to
    around 1.3~R$_{\odot}$ and then decreases up to 1.7~R$_{\odot}$.
    An increase in the line width with height is clearly visible for Fe~{\sc x} data.
    On 5 September (lower panels of Fig.~\ref{plotsmin11aug}), at PA 290 an increase is observed in the
    Fe~{\sc xiv} line width with height up to around 1.3~R$_{\odot}$ and then a
    decrease up to around 1.8~R$_{\odot}$. The Fe~{\sc x} data are not observed at this polar angle
     because of the low count rates.
    \begin{figure*}[th!]
    \centering
    \includegraphics[width=.32\textwidth,type=eps,ext=.eps,read=.eps]{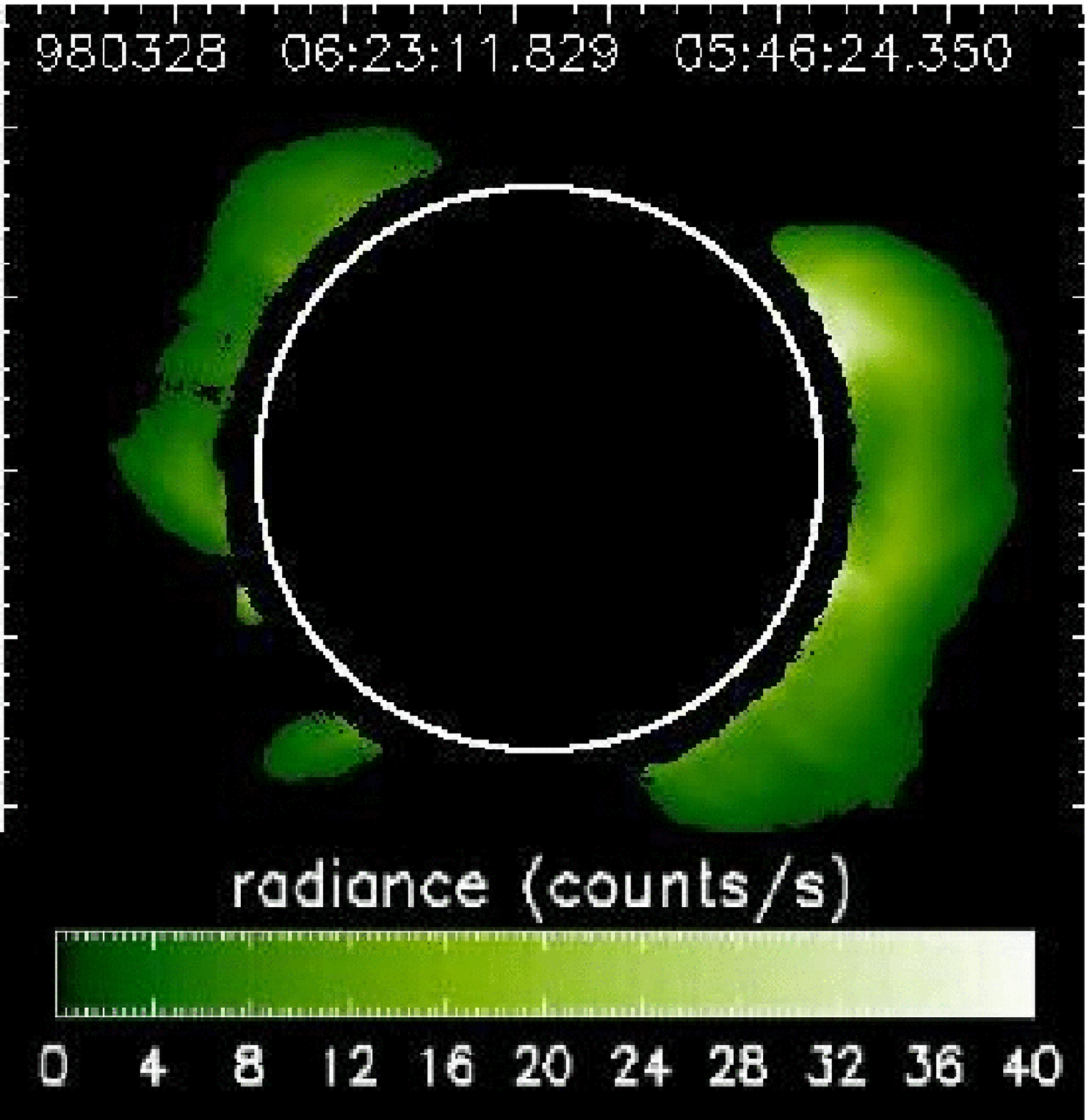}
    \includegraphics[width=.32\textwidth,type=eps,ext=.eps,read=.eps]{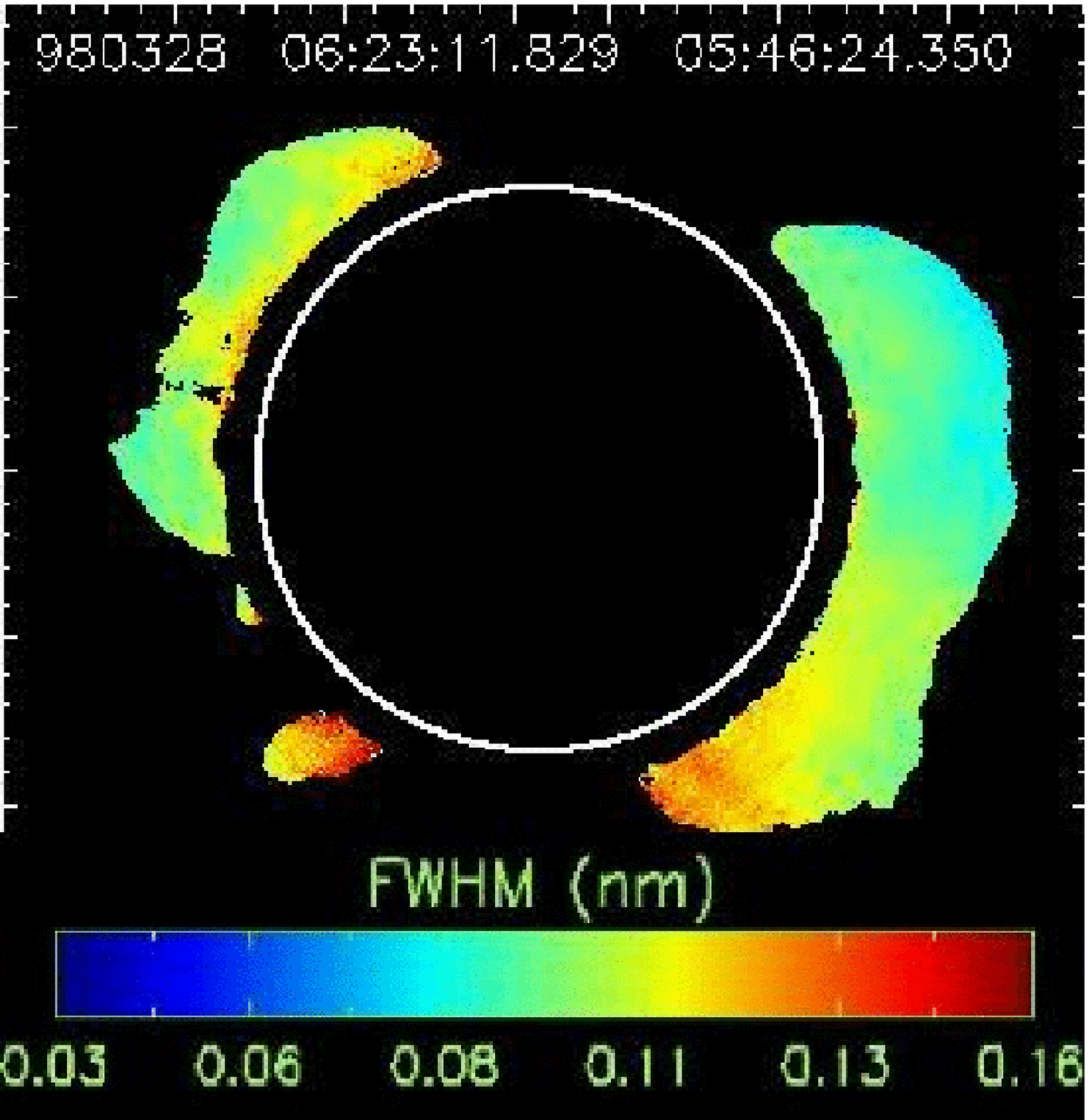}
    \includegraphics[width=.35\textwidth,type=eps,ext=.eps,read=.eps]{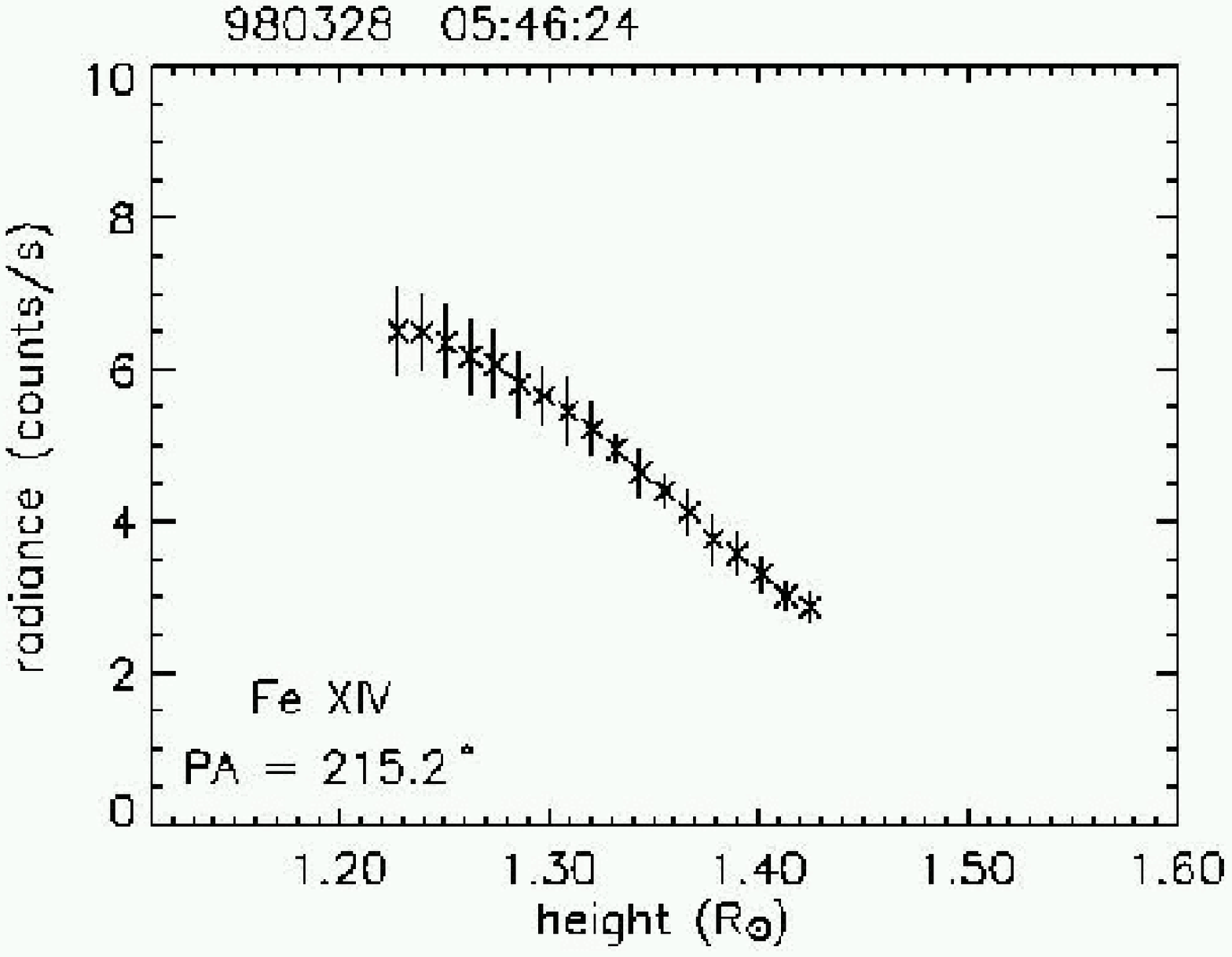}
    \includegraphics[width=.35\textwidth,type=eps,ext=.eps,read=.eps]{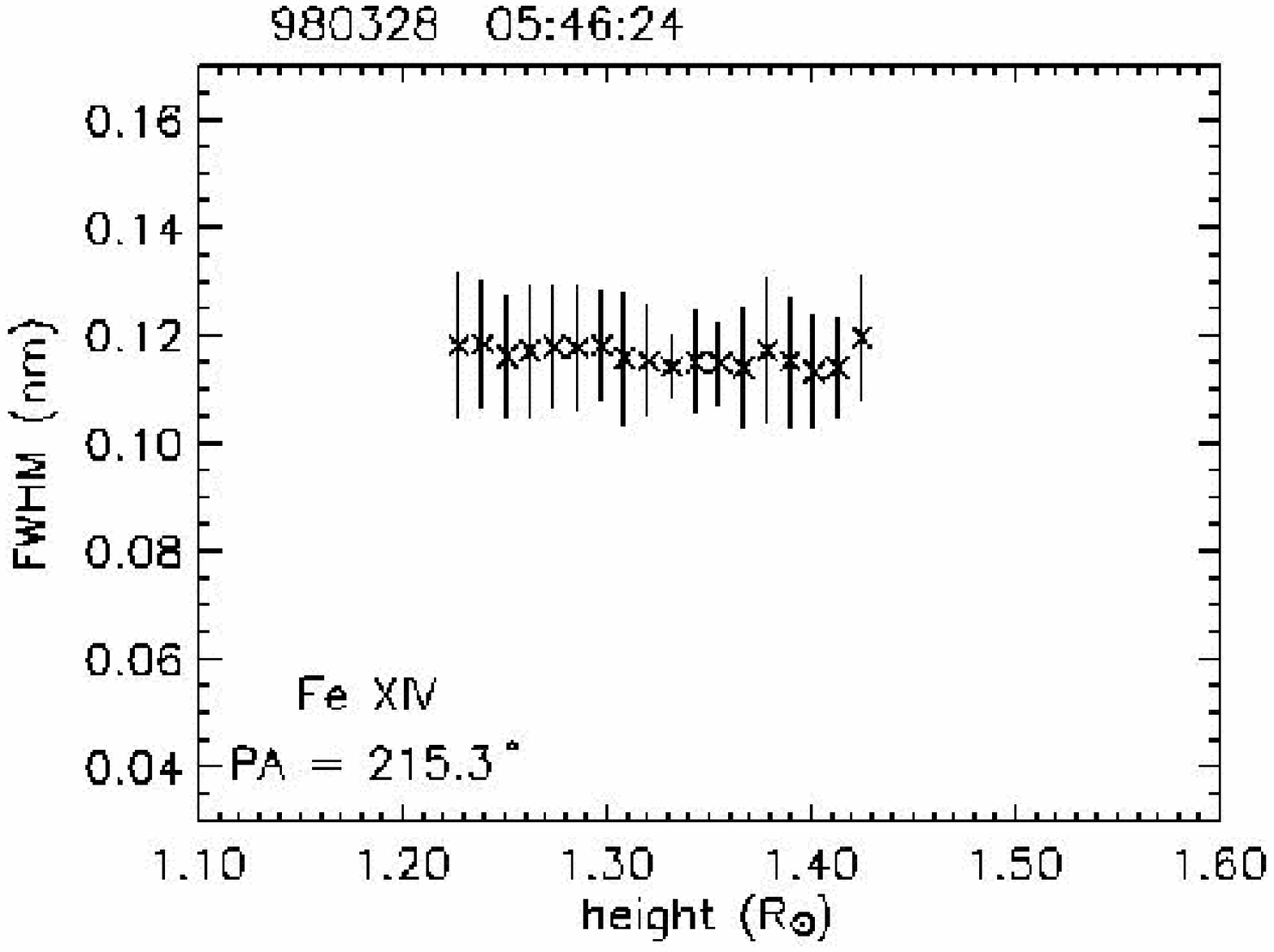}
    \includegraphics[width=.35\textwidth,type=eps,ext=.eps,read=.eps]{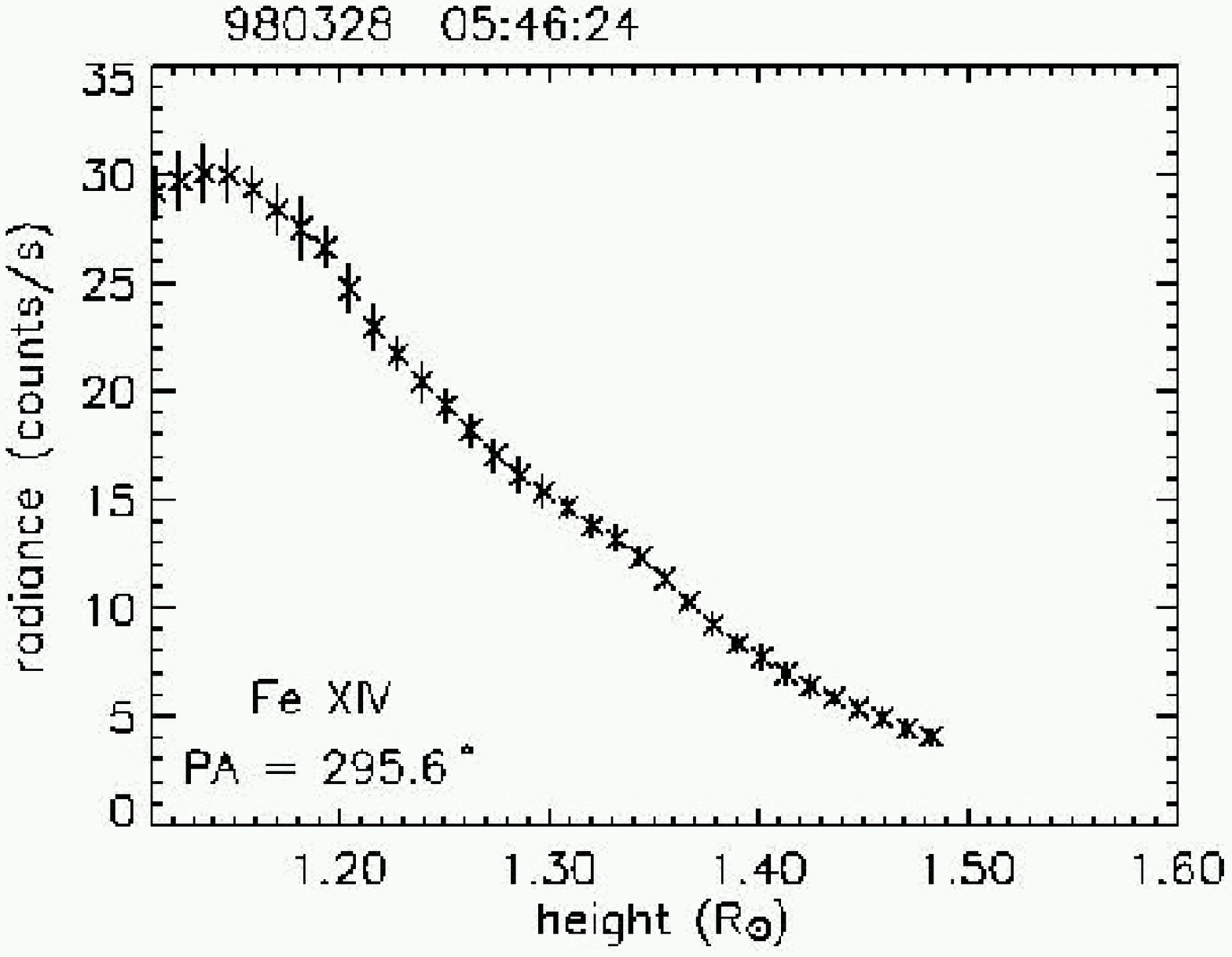}
    \includegraphics[width=.35\textwidth,type=eps,ext=.eps,read=.eps]{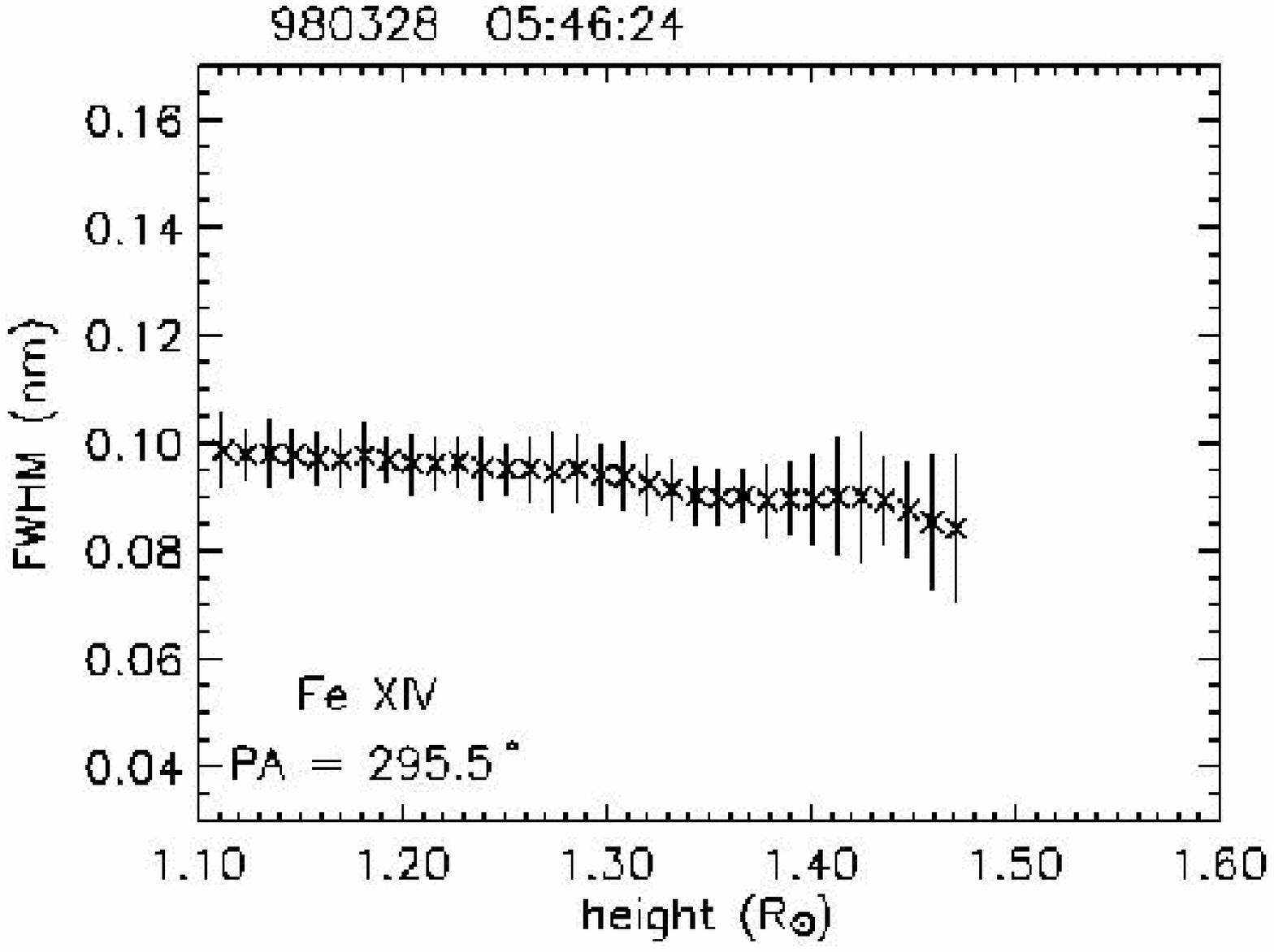}
    \caption{Fe~{\sc xiv} (upper panels) maps of radiance and FWHM
    for the data recorded on 28 of March 1998. The
    radial plots of the emission line parameters (middle and bottom panels)
    were taken at the indicated polar angles.}
    \label{plotsmax28march}
    \end{figure*}
    \begin{figure*}[th!]
    \centering
    \includegraphics[width=.34\textwidth, type=eps,ext=.eps,read=.eps]{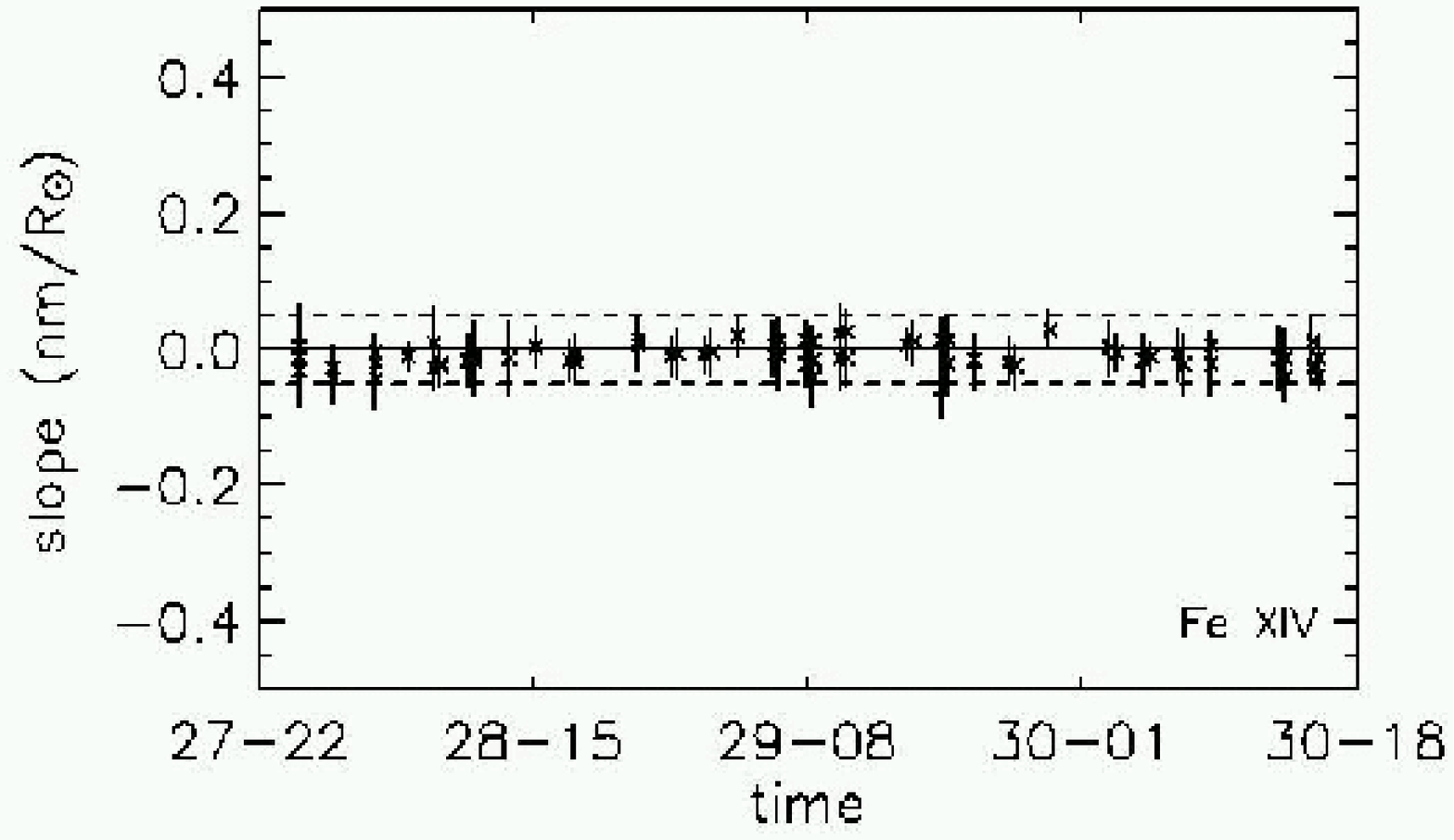}
    \includegraphics[width=.34\textwidth, type=eps,ext=.eps,read=.eps]{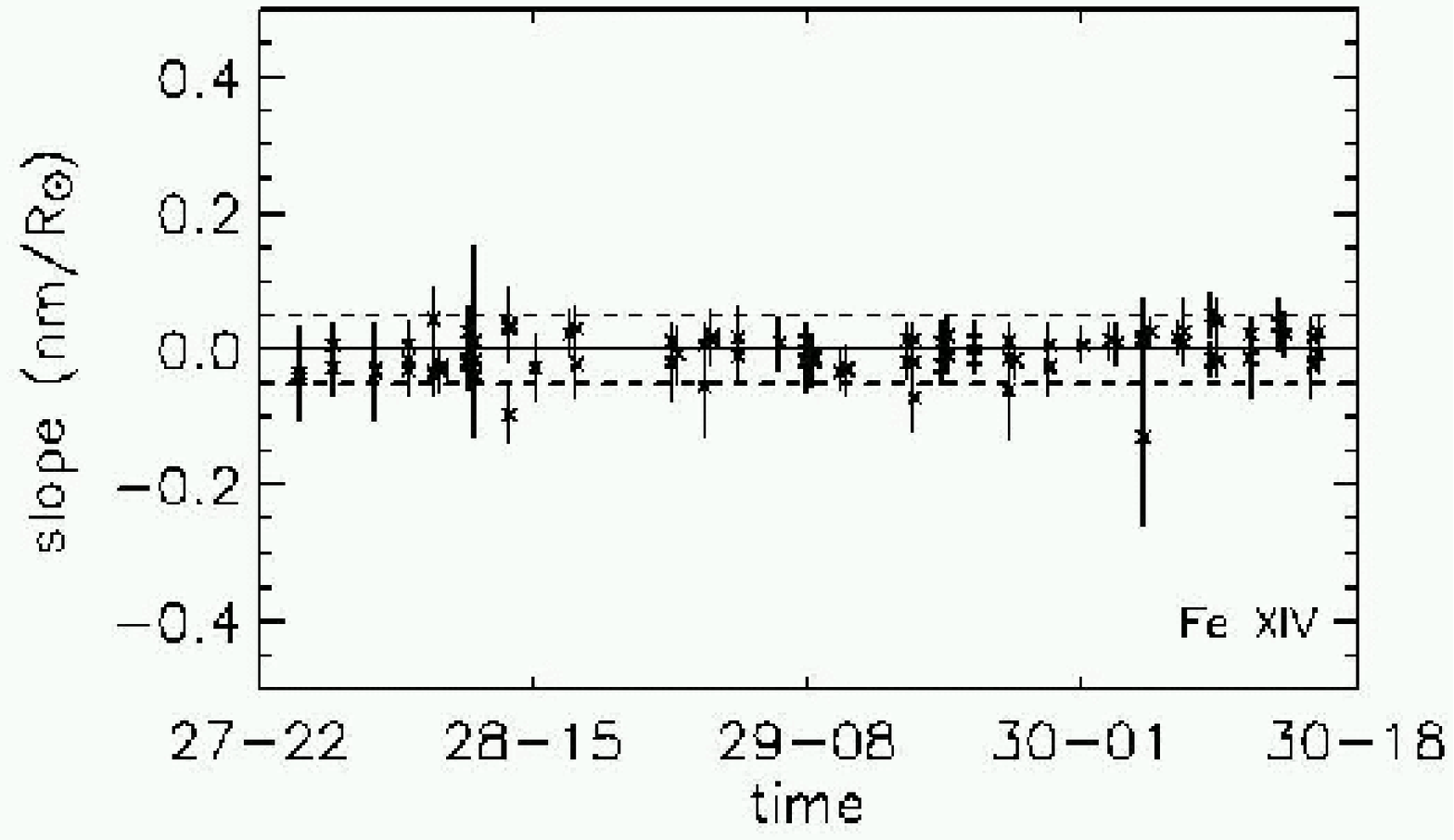}
    \includegraphics[width=.34\textwidth, type=eps,ext=.eps,read=.eps]{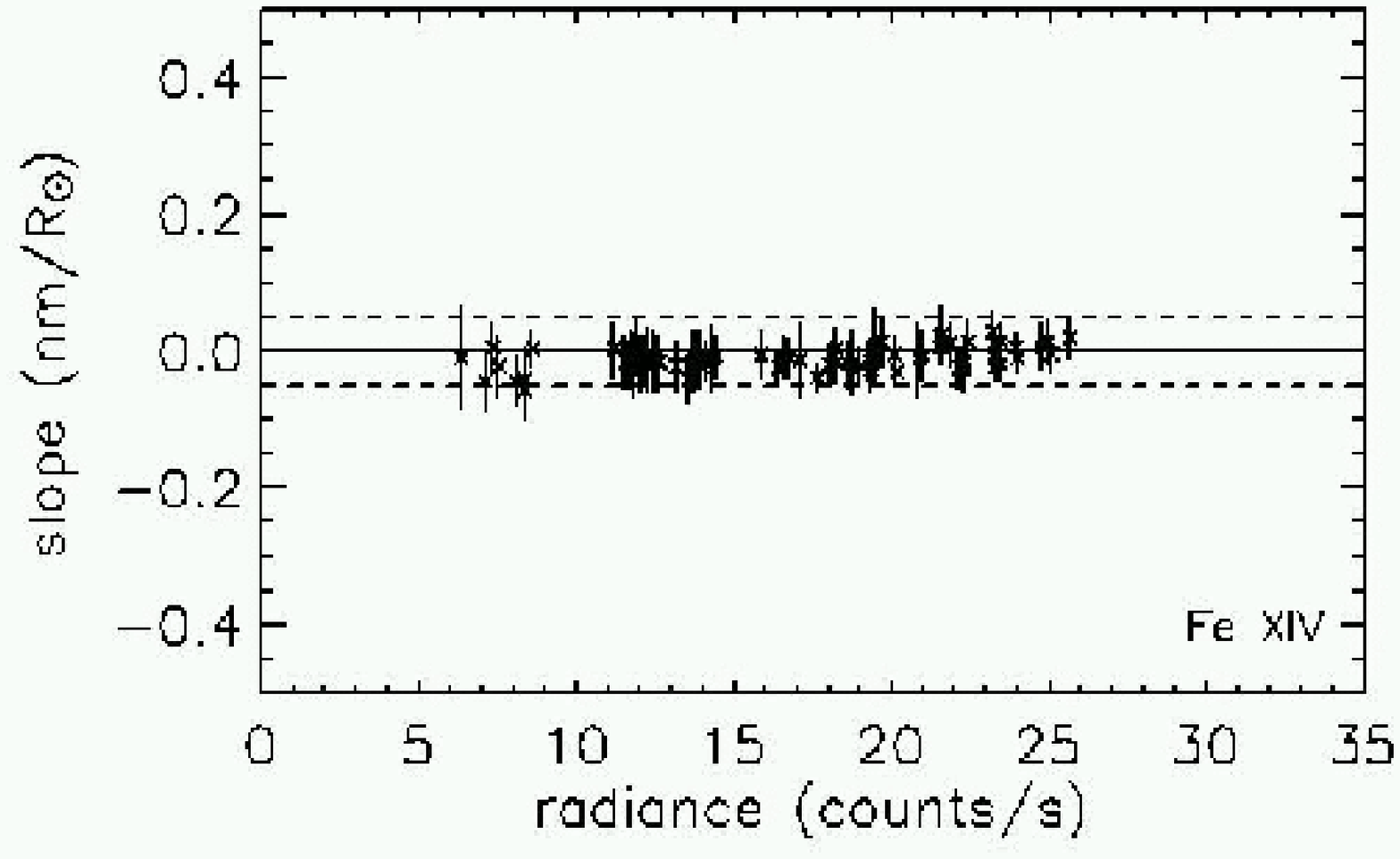}
    \includegraphics[width=.34\textwidth, type=eps,ext=.eps,read=.eps]{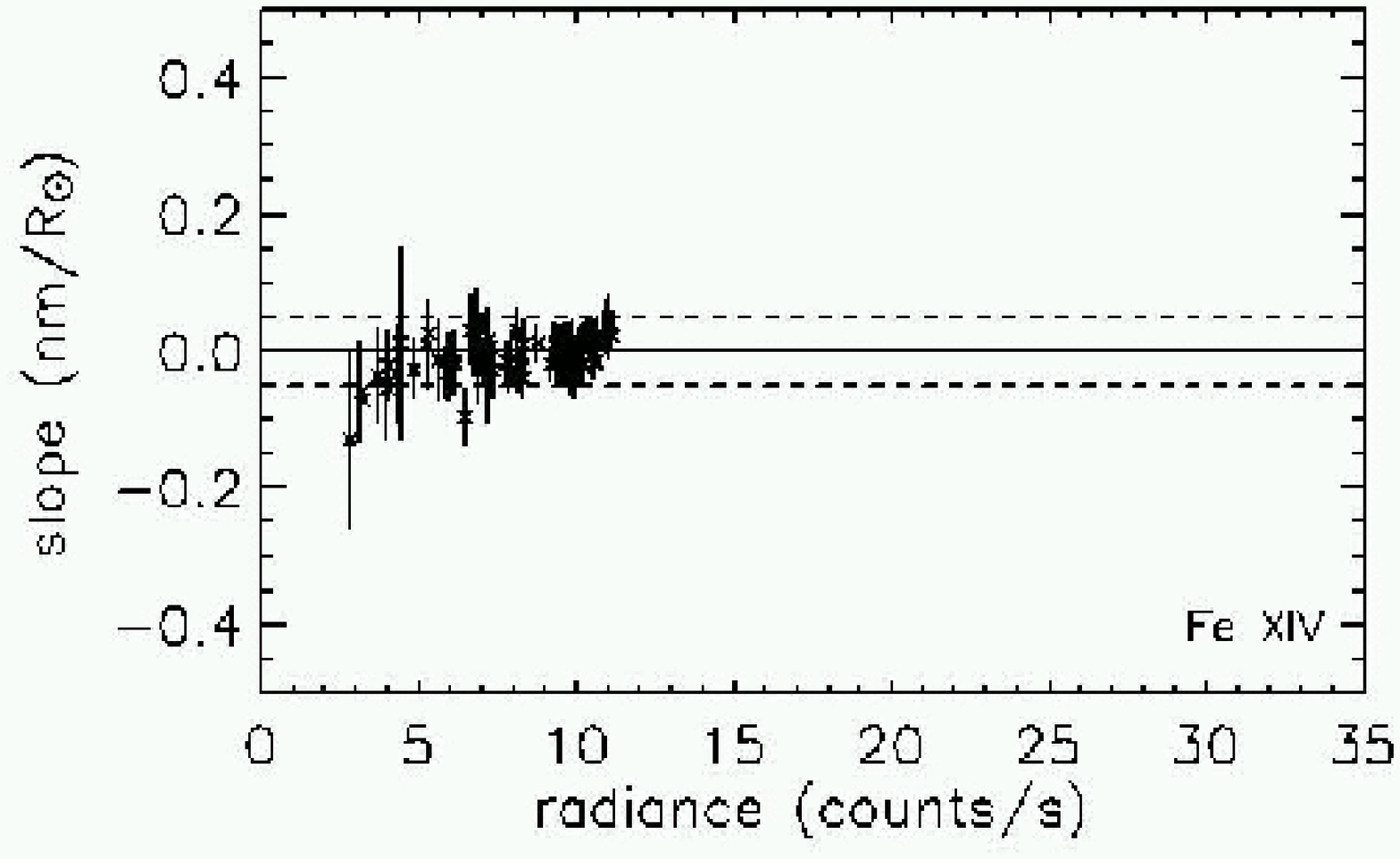}
    \caption{Upper panels: The variation in Fe~{\sc xiv} slope with time for the data in 1998.
    The upper-left panel shows the slopes for the data between
    1.1 and 1.3~R$_{\odot}$ and the upper-right panel, data between 1.3 and 1.5~R$_{\odot}$.
    The slopes are calculated for the data taken at the PA: 216,
    250, 270, 300. On the X-axis the date is represented in format dd-hh. Lower panels: The variation
    in Fe~{\sc xiv} slope with the radiance at 1.2~R$_{\odot}$ (lower-left panel)
    and 1.4~R$_{\odot}$ (lower-right panel) for the same data as in the above panels.}
    \label{fexivslope98}
    \end{figure*}

    \subsection{The FWHM dependence of height for green and red line data}
    To analyze the radial profile of the line widths, we make
    radial cuts at a given PA. We apply linear fits to
    the measured line widths and determine the slope. For each data
    set, we make cuts at 6 different polar angles. We plot the slopes of
    all data sets versus time. We excluded the points where the
    correlation coefficient is between -0.5 and 0.5. For Fe~{\sc x}
    data in 1996, we plot the slope in time for PA: 70, 90, 110,
    250, 270, 290 as shown in Fig.~\ref{fexslope96}, left panel.
    We see that, for the majority of the points, the slope is positive,
    i.e., the Fe~{\sc x} FWHM is increasing with radial distance throughout
    the entire time interval. Only in a few cases is the line width decreasing with radial distance.

    Inspection of the data leads us
    to the suspicion that the FWHM of Fe~{\sc x} may be affected by low count rates.
    To study this we plot the slope as
    function of line radiance for a constant radial distance of
    1.2~R$_{\odot}$. The result is shown in the right panel of Fig.~\ref{fexslope96}. We observe that the few negative
    slopes generally correspond to low radiances.
    For Fe~{\sc xiv} data, we first plot the points in the range 1.1
    to 1.3~R$_{\odot}$ (Fig.~\ref{fexivslope96}, upper left panel) and then 1.3 to 1.5~R$_{\odot}$
    (Fig.~\ref{fexivslope96}, upper right panel).
    We notice that, up to 1.3~R$_{\odot}$, the slope is generally
    positive; i.e., there is almost at all times an increase in the
    Fe~{\sc xiv} FWHM with height. For larger distances, however, the Fe~{\sc xiv} FWHM appears
    to decrease with height. The slope of Fe~{\sc xiv} FWHM versus
    radiance is shown in the lower panels of Fig.~\ref{fexivslope96}.

    \subsection{The line widths and the solar corona on the ascending phase of the solar cycle}

    From Fig.~\ref{plotsmax28march} we find that the radiance in the
    active region (AR) (lower left panel) is 5 times higher than the
    radiance in the non-active area (middle left panel). The steep
    decrease of the radiance with height at PA 215 is due to the fact
    that the feature is not radial and the cut just passes through a
    part of it. In both cases a slight decrease of the FWHM with
    distance is observed.

    For Fe~{\sc xiv} data in 1998 we plot the slope (as described in section 4.2) in time for PA: 216, 250, 270,
    300 as shown in Fig.~\ref{fexivslope98}, upper panels.
    We see that for the majority of the points the slope shows almost no trend, i.e. the width
    is almost constant with radial distance (see
    Fig.~\ref{fexivslope98}). The lower panels of
    Fig.~\ref{fexivslope98} shows the slope versus radiance.

    \section{Summary}

Line profiles observed with the LASCO-C1 instrument imply ion
temperatures much higher than the line formation temperature. This
means that, besides thermal motions, nonthermal effects, such as
turbulence, may be involved or else ions are heated to much higher
temperatures than electrons. As mentioned in the introduction,
there is debate regarding the variation in the line width with
height. Some authors have observed a decrease in the line width
with height above the solar limb and some found an increase. One
possible explanation for the increase in the line width with
height is the outward propagation of undamped Alfv\'{e}n waves and
for line width decrease with height is the resonant energy
conversion from Alfv\'{e}n to acoustic waves.
 From the data we have analyzed here we could see that, (1) in general,
we see different features in red and green line images, (2) the
red solar corona in 1996 was ``hotter'' than the green solar
corona in the same period. That means that the amount of
turbulence was higher in the cooler plasma or that there was
preferential heating of the Fe~{\sc x} ions, pointing to
ion-cyclotron heating. (3) For the data on 1996, the line width of
Fe~{\sc xiv} is roughly constant or increases with height up to
around 1.3~$R_{\odot}$ and then it decreases to around
1.5~$R_{\odot}$. The Fe~{\sc x} line width increases with height
up to around 1.3~$R_{\odot}$. For the data in 1998 the line width
of Fe~{\sc xiv} is roughly constant with height above the limb. No
Fe~{\sc x} data were available on that period. (4) The effective
temperatures are much higher than than those temperatures at which
Fe~{\sc xiv} and Fe~{\sc x} ions form (around 2~MK and 1~MK,
respectively). The excess is due to non-thermal motions (like
turbulences, wave motions, etc.), LOS effects, and/or ion
temperatures much higher than electron temperatures.

\begin{acknowledgements}
M.M. is thankful to Max Planck Institut f\"ur
Sonnensystemforschung for the facilities to carry out this work.
We are grateful to E. Marsch and R. Mecheri for productive
discussions. We would like to thank the SOHO/LASCO/EIT consortia
for providing the data and the software libraries. SOHO is a
project of international cooperation between ESA and NASA. We also
thank the anonymous referee for useful comments and suggestions.
\end{acknowledgements}

\end{document}